\documentclass[conference]{IEEEtran}
\IEEEoverridecommandlockouts
\usepackage{cite}
\usepackage{amsmath,amssymb,amsfonts}
\usepackage{bm}
\usepackage{subcaption}
\usepackage{adjustbox}
\usepackage{makecell}
\usepackage{multirow}
\usepackage{pgfplots}
\usepackage{colortbl}
\usepackage{tabularx}
\usepackage{tikz}
\usepackage{soul}
\usepackage{array}
\usepackage{hyperref}
\usepackage{caption}
\usepackage{ulem}
\usepackage{bbding}
\usepackage{pifont}
\usepackage{enumerate}
\usepackage{balance}
\usepackage{enumitem}
\usepackage{booktabs}
\usepackage{twemojis}
\usepackage{url}
\usetikzlibrary{patterns,shapes.geometric}
\usepackage[para,online,flushleft]{threeparttable}
\usepackage[ruled]{algorithm2e}
\usepgfplotslibrary{statistics,colorbrewer,colormaps}
\newcommand*\circled[1]{\tikz[baseline=(char.base)]{
            \node[shape=circle,draw,inner sep=1pt] (char) {#1};}}
\def\BibTeX{{\rm B\kern-.05em{\sc i\kern-.025em b}\kern-.08em
    T\kern-.1667em\lower.7ex\hbox{E}\kern-.125emX}}
\begin{document}

\title{Grounding Natural Language to SQL Translation with Data-Based Self-Explanations}

\author{\IEEEauthorblockN{Yuankai Fan\textsuperscript{1, 2}, Tonghui Ren\textsuperscript{1}, Can Huang\textsuperscript{1}, Zhenying He\textsuperscript{1}, X. Sean Wang\textsuperscript{1}}
\IEEEauthorblockA{
\textsuperscript{1}\textit{School of Computer Science, Fudan University}, Shanghai, China \\
\textsuperscript{2}\textit{Institute of Artificial Intelligence (TeleAI), China Telecom}, China \\ 
fanyuankai@fudan.edu.cn, \{thren22, huangcan22\}@m.fudan.edu.cn} \{zhenying, xywangCS\}@fudan.edu.cn 
}

\maketitle

\begin{abstract}

Natural Language Interfaces for Databases empower non-technical users to interact with data using natural language (NL). Advanced approaches, utilizing either neural sequence-to-sequence or more recent sophisticated large-scale language models, typically implement NL to SQL (NL2SQL) translation in an end-to-end fashion. However, like humans, these end-to-end translation models may not always generate the best SQL output on their first try. In this paper, we propose \textsc{Cyclesql}, an iterative framework designed for end-to-end translation models to autonomously generate the best output through self-evaluation. The main idea of \textsc{Cyclesql} is to introduce \textit{data-grounded NL explanations} of query results as self-provided feedback, and use the feedback to validate the correctness of the translation iteratively, hence improving the overall translation accuracy. Extensive experiments, including quantitative and qualitative evaluations, are conducted to study \textsc{Cyclesql} by applying it to seven existing translation models on five widely used benchmarks. The results show that 1) the feedback loop introduced in \textsc{Cyclesql} can consistently improve the performance of existing models, and in particular, by applying \textsc{Cyclesql} to \textsc{Resdsql}, obtains a translation accuracy of 82.0\% (+2.6\%) on the validation set, and 81.6\% (+3.2\%) on the test set of \textsc{Spider} benchmark; 2) the generated NL explanations can also provide insightful information for users, aiding in the comprehension of translation results and consequently enhancing the interpretability of NL2SQL translation\footnote{Our code is available at
\url{https://github.com/Kaimary/CycleSQL}.}.
\end{abstract}


\definecolor{pp}{rgb}{0.53, 0, 0.50}
\definecolor{rd}{rgb}{0.77, 0, 0}
\definecolor{pale}{rgb}{0.98, 0.91, 0.85}
\definecolor{sea}{rgb}{0.27, 0.63, 0.64}
\definecolor{dustyorange}{rgb}{0.96, 0.52, 0.43}

\section{Introduction}\label{intro}

Natural language interfaces for databases (NLIDBs) \cite{Androutsopoulos95, theory03, Practice22} democratize data exploration by allowing users to interact with databases using natural language (NL), breaking down barriers to information retrieval and data analysis. Consequently, the development of NLIDBs has garnered significant attention from both the data management and natural language processing (NLP) communities since the late 1970s \cite{ATIS90, GEO96, Zhong17, Yu18}.

In light of recent advancements in machine learning, the central focus of ongoing efforts to develop NLIDBs centers around elevating the accuracy of translating NL to SQL (NL2SQL) \cite{sigmod20, where20}. This objective is primarily accomplished via direct NL2SQL translation in an end-to-end manner, either by adopting sequence-to-sequence (Seq2seq) models trained on annotated data \cite{sqlnet17, IRNet19, Bogin19, RATSQL20, smbop21, LGESQL20, PICARD21, resdsql23} or, more recently, by harnessing large-scale language models (LLMs) that push the boundaries of the field even further within the last two years \cite{GPT4, DINSQL23, SQL-PaLM23, c323, DAILSQL23}.

\definecolor{indigo}{rgb}{0.43, 0.35, 0.83}
\definecolor{sea}{rgb}{0.27, 0.63, 0.64}
\definecolor{dustyorange}{rgb}{0.96, 0.52, 0.13}
\definecolor{bluegray}{rgb}{0.4, 0.6, 0.8}
\definecolor{bubblegum}{rgb}{0.99, 0.76, 0.8}
\pgfdeclareplotmark{mystar}{
    \node[star,star point ratio=2.25,minimum size=11pt, inner sep=0pt, draw=none, solid, fill=sea] {};}
\begin{figure}[t!]
    \centering
    \begin{tikzpicture}
    \begin{axis}[
        height=4.0cm,
        width=\linewidth,
        xlabel={Beam Size (Number of Chat Completion Choices)},
        ylabel={Translation Accuracy},
        xlabel style={at={(axis description cs:0.5,.05)},anchor=north,font=\small},
        ylabel style={at={(axis description cs:0.08,.5)},anchor=south,font=\small},
        axis line style=ultra thick,
        xmin=0.5, xmax=5.5,
        ymin=0.60, ymax=0.90,
        xtick={1,2,3,4,5},
        ytick={0.7,0.75,0.8,0.85},
        legend pos=south east,
        legend cell align={left},
        legend columns=2,
        legend style={
          nodes={scale=0.6, transform shape},
          very thin  
        },
        xmajorgrids=true,
        ymajorgrids=true,
        grid style=dashed,
        every axis plot/.append style={ultra thick}
    ]
    \addplot[
        indigo,
        mark=*,
        mark size=2pt,
    ]
    coordinates {
        (1,0.793)(2, 0.823)(3,0.841)(4,0.855)(5,0.862) 
    };
    \addplot[
        dustyorange,
        mark=triangle*,
        mark size=2pt,
    ]
    coordinates {
        (1,0.775)(2, 0.809)(3,0.824)(4,0.836)(5,0.844) 
    };
    \addplot[
        gray,
        mark=square*,
        mark size=2pt,
    ]
    coordinates {
        (1,0.695)(2, 0.762)(3,0.794)(4,0.812)(5,0.825) 
    };
    \addplot[
        bluegray,
        mark=diamond*,
        mark size=2pt,
    ]
    coordinates {
        (1,0.759)(2, 0.814)(3,0.836)(4,0.851)(5,0.859) 
    };
    \legend{\textsc{Picard (t5-3b)}, \textsc{Resdsql (t5-large)}, \textsc{Gpt-3.5-turbo}, \textsc{Dailsql (chatgpt)}}
    \end{axis}
    \end{tikzpicture}
    \vspace{-2mm}
    \captionsetup{font=small}
    \caption{Translation accuracy on \textsc{Spider} validation set with varied beam sizes (or chat completion choices). Accuracy for beam sizes (or chat completions) $>$ 1 is evaluated by matching any beam result.}
    \label{fig: model beam results}
    \vspace{-5mm}
\end{figure}
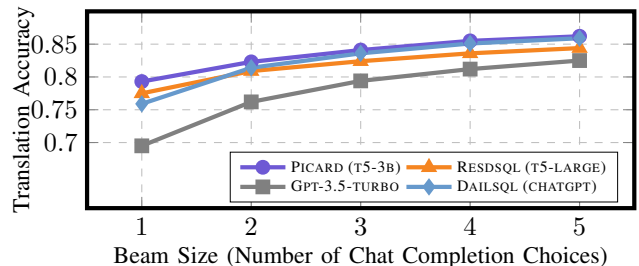

\noindent\textbf{Challenges.} While significant advancements in enhancing overall accuracy, \textit{current end-to-end models face persistent challenges in producing desired quality output during their initial attempt, owing to the treatment of language translation as a ``one-time deal''.} Fig.~\ref{fig: model beam results} shows the translation accuracy (defined as query execution result equivalence) on the \textsc{Spider} \cite{Yu18} benchmark, with varied beam outputs for two Seq2seq-based models\footnote{Most existing Seq2seq-based NL2SQL translation models utilize beam search decoding method to maintain a list of top-K best outcomes.} (i.e., \textsc{Picard} \cite{PICARD21}, and \textsc{Resdsql} \cite{resdsql23}) or diverse chat completion choices for two LLM-based models\footnote{LLMs often use a specific parameter to generate various responses. Refer to \url{https://platform.openai.com/docs/api-reference/chat/create\#chat-create-n}.} (i.e., \textsc{Gpt-3.5-turbo} and the state-of-the-art \textsc{Dailsql} model \cite{DAILSQL23}). As indicated by the plateauing accuracy (below 80\%) observed when the beam size or the number of chat completions is set to $1$, they may fail to generate best-quality translations in their initial attempts. However, expanding the search space through wider beams or more chat completions steadily improves accuracy, without necessitating modifications to the underlying model architectures. This example shows that end-to-end models may benefit from broader exploration options to enhance translation quality over successive attempts.

\newcommand{\coloredwave}[2]{
    \textcolor{#1}{\uwave{\textcolor{black}{#2}}}
}
\newcommand\boldpp[1]{\textcolor{pp}{\textbf{#1}}}
\begin{figure}[t]
 \begin{subfigure}[b]{1.0\linewidth}
 \centering
 \begin{adjustbox}{width=.45\linewidth}
    \begin{tabular}{c c c c}
        \hline
        \rowcolor{pale} \textbf{aid} & \textbf{flno} & \textbf{origin} & \textbf{destination} \\ \hline
        9 & 2 & Los Angeles & Tokyo \\ 
        3 & 7 & Los Angeles & Sydney \\ 
        3 & 13 & Los Angeles & Chicago   \\ 
        10 & 68 & Chicago & New York \\
        9 & 76 & Chicago & Los Angeles \\
        7 & 33 & Los Angeles & Honolulu  \\ 
        5 & 34 & Los Angeles & Honolulu  \\
        1 & 99 & Los Angeles & Washington D.C. \\ 
        2 & 346 & Los Angeles & Dallas \\ 
        6 & 387 & Los Angeles & Boston \\  \hline 
    \end{tabular}
\end{adjustbox}\hspace{3mm}
 \begin{adjustbox}{width=.47\linewidth}
    \begin{tabular}{c c c c}
        \hline
        \rowcolor{pale} \textbf{aid} & \textbf{name} & \textbf{distance} \\ \hline
        1 & Boeing 747-400 & 8430 \\ 
        2 & Boeing 737-800 & 3383 \\ 
        3 & Airbus A340-300 & 7120   \\ 
        4 & British Aerospace Jetstream 41 & 1502 \\
        5 & Embraer ERJ-145 & 1530  \\
        6 & SAAB 340 & 2128 \\ 
        7 & Piper Archer III & 520 \\ 
        8 & Tupolev 154 & 4103 \\ 
        9 & Lockheed L1011 & 6900 \\
        10 & Boeing 757-300 & 4010 \\\hline 
    \end{tabular}
    \end{adjustbox}
 \label{fig:(a)spider query}
 \end{subfigure}
 \begin{subfigure}[b]{1.0\linewidth}
    \centering
    \includegraphics[width=\linewidth]{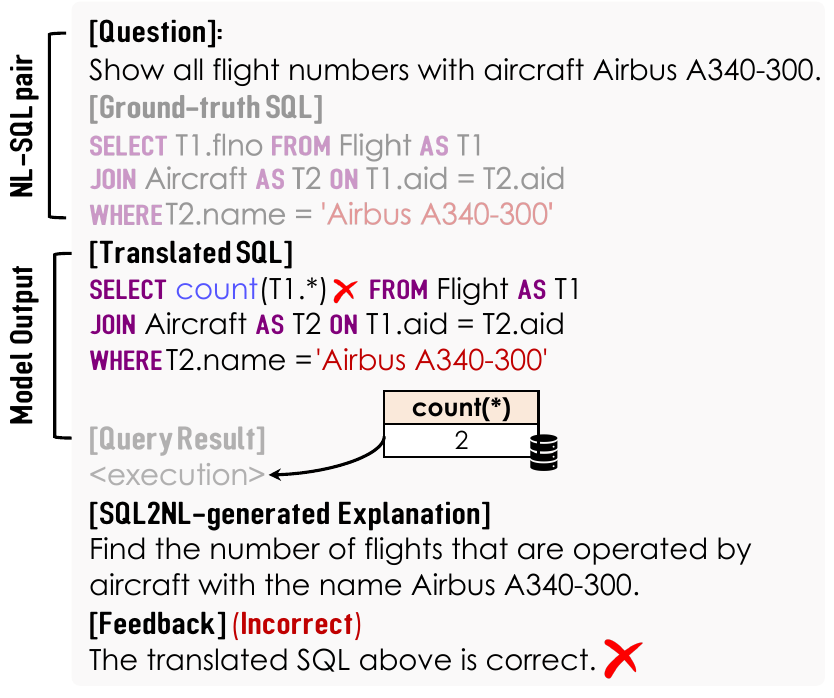}
 \end{subfigure}
 \vspace{-7mm}
 \captionsetup{font=small}
 \caption{{A simplified example from the \textsc{Spider} benchmark.}}
 \label{fig: spider query}
 \vspace{-5mm}
\end{figure}


Inspired by the feedback mechanisms \cite{feedback16, feedback20} used in modern recommendation systems and iterative refinement methods \cite{ReidN22, self-refiner23} introduced in LLMs, we present \textsc{\textbf{Cyclesql}}, an iterative framework built upon \textit{self-provided feedback} to enhance translation accuracy of existing end-to-end models. Diverging from the traditional end-to-end translation paradigm, we introduce data-grounded NL explanations of query results as a form of internal feedback to create a self-contained feedback loop within the end-to-end process, facilitating iterative self-evaluation of translation correctness.

\noindent\textbf{\twemoji[height=1.0em]{thinking face}SQL2NL back-translation for Self-Explanations?} A natural way to generate self-provided feedback (explanations) in NL2SQL translation may involve its reverse process, namely SQL2NL. This entails establishing an NL-to-SQL-to-NL translation lifecycle, a concept analogous to back-translation \cite{BT16, EdunovOAG18}. Several studies have explored this technique \cite{DanikenDABCS22, sciencebenchmark23, self-debug23, chess24} to refine initial SQL outputs, but we posit that \textit{solely using this ``simple'' back-translation to generate feedback may be limited, as it lacks additional contextual information beyond the NL and SQL components}. Consider an intuitive example in Fig.~\ref{fig: spider query}, which shows an NL query alongside a translation from an existing model. By simply using the SQL2NL technique, however, incorrect ``positive'' feedback is generated: the meaning of the explanation seems to align with the initial NL query, whereas the underlying SQL query is deemed incorrect {(i.e., an incorrect aggregation function is used in the {\fontfamily{pcr}\fontsize{9}{10}\selectfont\textcolor{pp}{\textbf{SELECT}}} clause)}. Thus, such a back-translated explanation may not adequately serve as valid feedback for the NL2SQL process.

\noindent 

\noindent\textbf{Our Methodology.} We propose a plug-and-play framework that facilitates seamless integration with existing end-to-end NL2SQL models. By integrating with a conventional NL2SQL process, \textsc{Cyclesql} \circled{1} first rewrites the translated SQL query to retrieve provenance information from the underlying database for a specific query result. \circled{2} Following this, the provenance is enriched by annotating the semantics of the translated SQL query to its different parts. Next, \circled{3} a data-grounded NL explanation is interpreted from the enriched provenance to rationalize the query result by leveraging text generation techniques \cite{Deutch17, Deutch20, GAR23, self-refiner23}. Here, \textit{given that the NL explanation integrates both data-level (provenance) and operation-level (query) semantics, they possess significant potential to function as self-provided feedback for various NL2SQL models.} Finally, \circled{4} the generated explanation is used to validate the correctness of the underlying translation iteratively, ultimately yielding a more reliable translation outcome.

To assess the effectiveness of \textsc{Cyclesql}, we conduct a comprehensive experimental evaluation on the widely-used benchmark \textsc{Spider}, and its three variants, namely \textsc{Spider-realistic} \cite{spider-realistic21}, \textsc{Spider-syn} \cite{spider-syn21}, and \textsc{Spider-dk} \cite{spider-dk21}, {as well as \textsc{ScienceBenchmark} \cite{sciencebenchmark23}}, by applying \textsc{Cyclesql} to seven contemporary end-to-end models. The results show that \textsc{Cyclesql} can consistently enhance the translation accuracy of all the models. Notably, by applying \textsc{Cyclesql} to \textsc{Resdsql} (with \textsc{T5-3b} scale), \textsc{Cyclesql} obtains a translation accuracy of 82.0\% on the validation set and 81.6\% on the test set of \textsc{Spider} benchmark, achieving best-reported result among the leading Seq2seq-based models on \textsc{Spider} leaderboard\footnote{\url{https://yale-lily.github.io/spider}}. Moreover, a qualitative evaluation, including a case study and user study, is conducted to show that the explanations generated by \textsc{Cyclesql} can also greatly improve user interpretability in the black-box NL2SQL process.

\noindent\textbf{Contributions.} We make the following three contributions:
\begin{enumerate}[leftmargin=*, label={(\arabic*)}]
 \item\textbf{Feedback Loop in NL2SQL.} We propose, \textsc{Cyclesql}, a plug-and-play framework to establish a self-contained feedback loop within the NL2SQL process. \textsc{Cyclesql} employs data-grounded NL explanations as reliable feedback to iteratively validate the correctness of the translation, thereby enhancing the overall accuracy.
  
 \item\textbf{Rich Explanations for NL2SQL.} The NL explanations generated by \textsc{Cyclesql}, which incorporate not only the semantics of the query surface but also the semantics from the data instance, provide insightful information for users to understand the black-box NL2SQL translation process.

 \item\textbf{Quantitative and Qualitative Evaluations.} {We evaluate \textsc{Cyclesql} on five public benchmarks with seven NL2SQL models}, demonstrating its substantial impact on performance improvements. Furthermore, a qualitative evaluation is conducted to gauge the utility of the generated NL explanations in enhancing user interpretability.
\end{enumerate}


\begin{figure*}[ht]
    \centering
    \includegraphics[width=\linewidth]{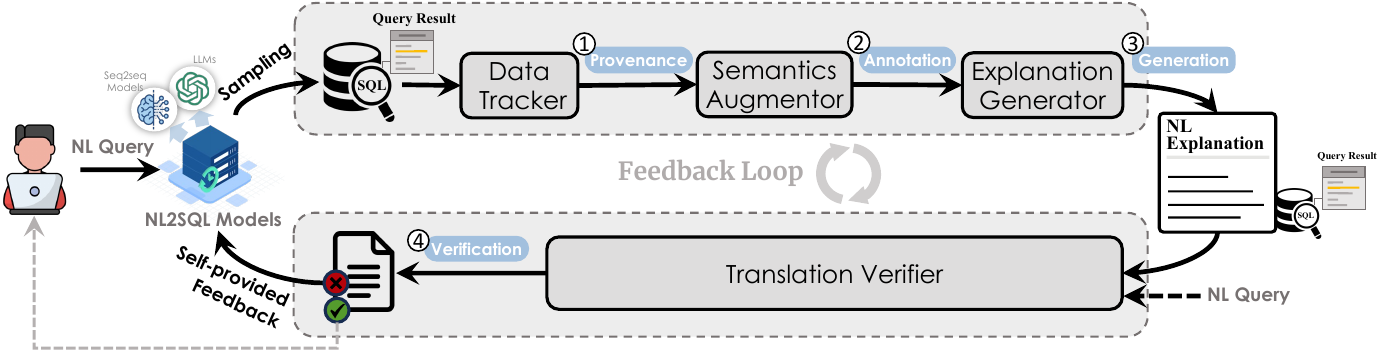}
    \vspace{-7mm}
    \captionsetup{font=small}
    \caption{Overview framework of \textsc{Cyclesql}}
    \label{fig: overview}
    \vspace{-6mm}
\end{figure*}

\section{Preliminaries}\label{section 2}

\def\mathbi#1{\textbf{\em #1}}
\subsection{NL2SQL models}
\noindent \textbf{Seq2seq-based NL2SQL Models.} A Seq2seq-based NL2SQL model typically adheres to the encoder-decoder learning framework \cite{seq2seq14} to translate NL queries to the corresponding SQL queries. Given an input NL query $X = \{x_{1},x_{2},\cdots,x_{n}\}$ and a database schema $S=\left\langle T, C\right\rangle$ that consists of a set of tables $T$ and a set of corresponding columns $C$, the model uses an encoder to compute a contextual representation \textbf{c} by jointly embedding the NL query $X$ with schema $S$. Afterward, an autoregressive decoder is used to calculate a distribution $P(Y\mid\textbf{c})$ over the SQL programs $Y=(y_{1},\cdots,y_{\mid Y\mid})$. 


\noindent\textbf{LLMs as NL2SQL Models.} In light of recent advancements in LLMs, current research \cite{nitarshan22, aiwei23} endeavors to employ LLMs as NL2SQL models without fine-tuning. Given an NL query $X$ and a prompt $P$ as input, an LLM can be utilized to generate the SQL query $Y$. Depending on the prompting technique utilized, such as zero-shot, few-shot prompting, or in-context learning, the prompt $P$ may include instructions \cite{aiwei23}, demonstrations \cite{SQL-PaLM23, DAILSQL23} or reasoning chains \cite{DINSQL23, Interleaving23}.

\subsection{Provenance} 
Provenance elucidates the origin and history of the data throughout its lifecycle. Research has studied provenance semantics for various query classes in databases \cite{Cheney09, Green07}. The most common forms of database provenance describe the relationships between the data in the source and its corresponding output \cite{Provenance09}. This includes explaining where the output data originated in the input (commonly known as where-provenance) \cite{where-provenance90}, revealing inputs that explain why an output record was produced (referred to as why-provenance) \cite{why-provenance00, why-where-provenance01}, or providing a detailed account of how an output record was generated (referred to as how-provenance) \cite{how-provenance07}. This work adopts \textit{why-provenance}, which functions as evidence within a query, comprising a subset of tuples that guarantees the presence of a specific tuple in the output.

 Formally, given a query $q$ having relations $\mathbb{R} = \{R_{j_{1}}, \cdots, R_{j_{p}}\}$, the why-provenance $\mathcal{P}(q, D)$ for database $D$ and the query $q$ to be a subset of $R_{j_{1}} \times \cdots \times R_{j_{p}}$, where a provenance model determines which tuples from the cross product belong to $\mathcal{P}(q, D)$. For a tuple $t \in q(D)$, the provenance $\mathcal{P}(q, D, t)$ is the subset of the provenance that contributes to $t$ (decided by the provenance model).

\section{\textsc{Cyclesql} Overview}\label{section 3}
This section first provides a formal definition of the NL explanation we defined and presents the overview of \textsc{Cyclesql}.

\subsection{Natural Language Explanation} \label{sec 3.1}
Explanations of database query answers can differ across various dimensions, encompassing clarifying unexpected query outcomes \cite{missing20}, explaining the rationale behind generating intriguing results from a query \cite{CaJaDE21}, and justifying how an answer aligns with the query criteria \cite{Deutch17, Deutch20}. Each of these dimensions serves a distinct purpose in query explanation.

In this work, we focus on the explanation for why an answer qualifies a given query and provide the following definition,


\noindent\textsc{Definition} 1. \textbf{(Explanation)} \textit{Given an NL query $X$ that can be parsed against the database $D$ to retrieve a result set R, an} explanation \textit{$e_{r}$ takes the form of the NL expression that explains the intermediate reasoning steps behind querying $X$ on database $D$ to obtain a representative result $r \in R$.}

\noindent\textsc{Definition} 2. \textbf{(Intermediate Reasoning Step)} \textit{An} intermediate reasoning step \textit{refers to a cognitive process that involves logical operations or transformations on the instances over database $D$, taking place between the NL query $X$ and the query result $r$. Each} reasoning step \textit{contributes to refining the understanding of $X$ and progressively approaching $r$.}


\noindent\textbf{Example 1.} The following shows an NL explanation generated by \textsc{Cyclesql} (details of the generation process will be explained in late sections), which mostly includes two intermediate reasoning steps specifying the details of the reasoning process from the initial NL query to the query result based on the translated SQL query in the example shown in Fig.~\ref{fig: spider query}.

\vspace{-3mm}
\begin{figure}[ht]
  \begin{tikzpicture}
  \node (table) [inner sep=0pt] {
  \begin{adjustbox}{width=\linewidth}
    \begin{tabular}{l} 
     \\ [-2.0ex]
     \makecell[l]{\textbf{Explanation:} The query returns a result with one column of aggregation  \\
     type (count) and one row. That is, \ul{for flights with aircraft Airbus A340-300,} \\
     \hspace{4.95cm} (\textit{\textbf{intermediate reasoning step}} \ding{202}) \\
    \ul{there are 2 flights in total.} \\\hspace{0.050cm} (\textit{\textbf{intermediate reasoning step}} \ding{203})} \\
     [-1.8ex] \\
  \end{tabular}
  \end{adjustbox}
  };
  \draw[line width=0.50mm, rounded corners=.5em](table.north west) rectangle (table.south east);
  \end{tikzpicture}
  \vspace{-9mm}
\end{figure}

\subsection{Feedback Loop in \textsc{Cyclesql}}\label{sec 3.2}
 A high-level view of \textsc{Cyclesql} is presented in Fig.~\ref{fig: overview}. When seamlessly integrated into the end-to-end NL2SQL process of a designated translation model, the fundamental operational sequence of \textsc{Cyclesql} unfolds as follows:

 \begin{enumerate}[leftmargin=*, label={(\arabic*)}]
  \item \textsc{Cyclesql} first tracks the provenance information of the to-explained query result to retrieve data-level information from the underlying database.
  \item To align with user intent, \textsc{Cyclesql} further enhances the provenance information by associating it with operation-level semantics derived from the translated SQL query.
  \item An NL explanation of the query result is then interpreted by \textsc{Cyclesql} based on the enriched provenance.
  \item The NL explanation is subsequently used to verify the correctness of the underlying translation, iterating through the above steps until a validated translation is achieved.
 \end{enumerate}

\noindent Among these, the ``hybrid'' semantics that joins data-level provenance and operation-level query semantics is unique to our setup, and we found that the semantics are crucial for establishing a reliable feedback loop for NL2SQL.

 \noindent\textbf{Data Tracking.} In the initial stage (illustrated in Fig.~\ref{fig: overview}-\circled{1}), \textsc{Cyclesql} entails tracing the lineage of the query result obtained from the execution of a SQL query sampled from an underlying translation model, intending to gather provenance information. In \textsc{Cyclesql}, the tracing process is implemented by rewriting the sampled SQL query to retrieve all relevant data-level details from the underlying database. 
 
 \noindent For example, regarding the query result shown in Fig.~\ref{fig: spider query}, {its provenance corresponds to tuples $F2$, $F3$, where the column ``aid'' is equal to $3$ in ``Flight'' table, as below,}

 \vspace{-3mm}
 \begin{figure}[h]
 \centering
 \begin{adjustbox}{width=.9\linewidth}
    \begin{tabular}{|c | c | c | c | c | c|}
        \hline
        \rowcolor{pale} \textbf{\textcolor{gray}{tupleID}} & \textbf{aid} & \textbf{name} & \textbf{flno} & \textbf{origin} & \textbf{destination} \\ \hline
        \textcolor{gray}{F2} & 3 & Airbus A340-300 & 7 & Los Angeles & Sydney \\ \hline
        \textcolor{gray}{F3} & 3 & Airbus A340-300 & 13 & Los Angeles & Chicago \\ \hline
    \end{tabular}
    \end{adjustbox}
    \vspace{-3mm}
 \end{figure}
 
 
 \noindent\textbf{Semantics Enrichment.}\label{section: annotation} In the following stage (depicted in Fig.~\ref{fig: overview}-\circled{2}), \textsc{Cyclesql} further enriches the provenance information by integrating the operation-level semantics from the translated SQL query into its different parts. Since the translated SQL query captures the semantics from the initial NL query, this integration enables the reflection of user query semantics across different data segments of provenance. \textsc{Cyclesql} achieves this by decomposing the SQL query into a set of query units and associating the semantics of each query unit with relevant parts of data provenance.
 
 \noindent Continuing the example in Fig.~\ref{fig: spider query}, the column ``name'' in the provenance table is enriched with the filtering semantics from the {\fontfamily{pcr}\fontsize{9}{10}\selectfont \textcolor{pp}{\textbf{WHERE}}} operation, and the whole table is associated with the {\fontfamily{pcr}\fontsize{9}{10}\selectfont\textcolor{blue}{count}} semantics from {\fontfamily{pcr}\fontsize{9}{10}\selectfont \textcolor{pp}{\textbf{SELECT}}} to attempt to reflect the ``\textit{all flight numbers}'' semantics in the original NL query.

 \noindent\textbf{Explanation Generation.} In the subsequent stage (depicted in Fig.~\ref{fig: overview}-\circled{3}), \textsc{Cyclesql} interprets the enriched provenance information into an NL expression, generating a detailed explanation for the given query result. Inspired by the success of text generation techniques, \textsc{Cyclesql} utilizes a simple but effective rule-based method to synthesize the NL explanation mechanically from the given enriched provenance information.
 
 
 \noindent For instance, the NL explanation produced by \textsc{Cyclesql} for query result in Fig.~\ref{fig: spider query} can be seen in Example $1$ in Section \ref{sec 3.1}, which interprets the meaning of number $2$.

 \noindent\textbf{Translation Verification.} In the final stage (shown in Fig.~\ref{fig: overview}-\circled{4}), \textsc{Cyclesql} employs the generated explanation as self-provided feedback to verify the accuracy of the underlying NL2SQL translation. To accomplish this, \textsc{Cyclesql} formulates the validation process as a \textit{natural language inference} (NLI) task and uses a deep-learning model to determine whether the NL explanation entails the original NL query, thereby validating the correctness of the translation. \textsc{Cyclesql} iterates through the above stages until validation succeeds, ultimately leading to an improved translation outcome.
 
 \noindent Continuing the example in Fig.~\ref{fig: spider query}, {\textsc{Cyclesql} may recognize that expressions such as ``\textit{there are 2 flights in total}'' in the explanation are not aligned with the ``\textit{all flight numbers}'' semantics of the initial NL query,} thereby the translated SQL query is validated as incorrect. In response, \textsc{Cyclesql} solicits another sampled SQL query from the underlying translation model, repeating this process until the validation succeeds.
 
\section{Methodologies}\label{section 4}

 \subsection{Provenance via Query Rewriting}\label{sec 4.1}
To explain an individual query result, the essential element is detailed provenance information. This information is akin to the \textit{provenance table} introduced in \cite{CaJaDE21}, serving as an immediate form of explanation that describes how a query result is derived from the data stored in the underlying database. Acknowledging the relational semantics in SQL, wherein each operator generates its result tuple by tuple from its operand tables \cite{lineage00}, we employ the \textit{query inversion} methods \cite{GlavicMA13, RaniGG15} to capture provenance information using query rewriting. In this way, \textsc{Cyclesql} inverts transformations performed by a given SQL query to determine the source tuples that have contributed towards the result set.

Specifically, given an executed SQL query with a specific to-explained query result, \textsc{Cyclesql} employs the following three heuristic rules to automatically rewrite the SQL into a query that can be executed to capture provenance.

\begin{itemize}[leftmargin=*]
 \item\textit{Rule 1 (Result Transformation Rule).} Intuitively, the query result may specify the particular column and value that satisfies the query criteria. Hence, the initial step in the rewriting process is to translate the query result into a {\fontfamily{pcr}\fontsize{9}{10}\selectfont \textcolor{pp}{\textbf{WHERE}}} clause to explicitly specify the result-specific condition and incorporate it back to the original query.
 

 \noindent Note that in some cases, a query result is not associated with a specific column (e.g., when referring to all columns in the referenced table using the asterisk symbol (*)). In such cases, the application of this rule may be skipped.

 \item\textit{Rule 2 (Projection Enhancement Rule).} We observe that every column referenced in a SQL query (irrespective of the associated operator) provides valuable hints for provenance. Therefore, we extract all columns used in the original SQL query and incorporate them as additional projection columns. Besides that, we also include the primary keys of those referenced tables as part of the projection columns.

 \noindent Acknowledge that retrieving all columns in the referenced tables for provenance may be more straightforward. However, introducing this rule can aid in crafting a more concise explanation using more relevant provenance information. 
 \item\textit{Rule 3 (Aggregation Deconstructive Rule).} Since aggregation functions (and the {\fontfamily{pcr}\fontsize{9}{10}\selectfont \textcolor{pp}{\textbf{GROUP BY}}} clause) may collapse input data rows, they inherently conceal the ability to track the provenance. Therefore, we invert the potential transformations by simply eliminating aggregation functions (and {\fontfamily{pcr}\fontsize{9}{10}\selectfont \textcolor{pp}{\textbf{GROUP BY}}} clause) to unravel the lineage of the data\footnote{Note that these operations are temporarily excluded for query rewriting purpose. They will be revisited in the following phase for query semantic annotation. More details can be found in the later section.}.
 \end{itemize}

 \begin{figure}[ht]
 \vspace{-3mm}
 \centering
 \includegraphics[width=\linewidth]{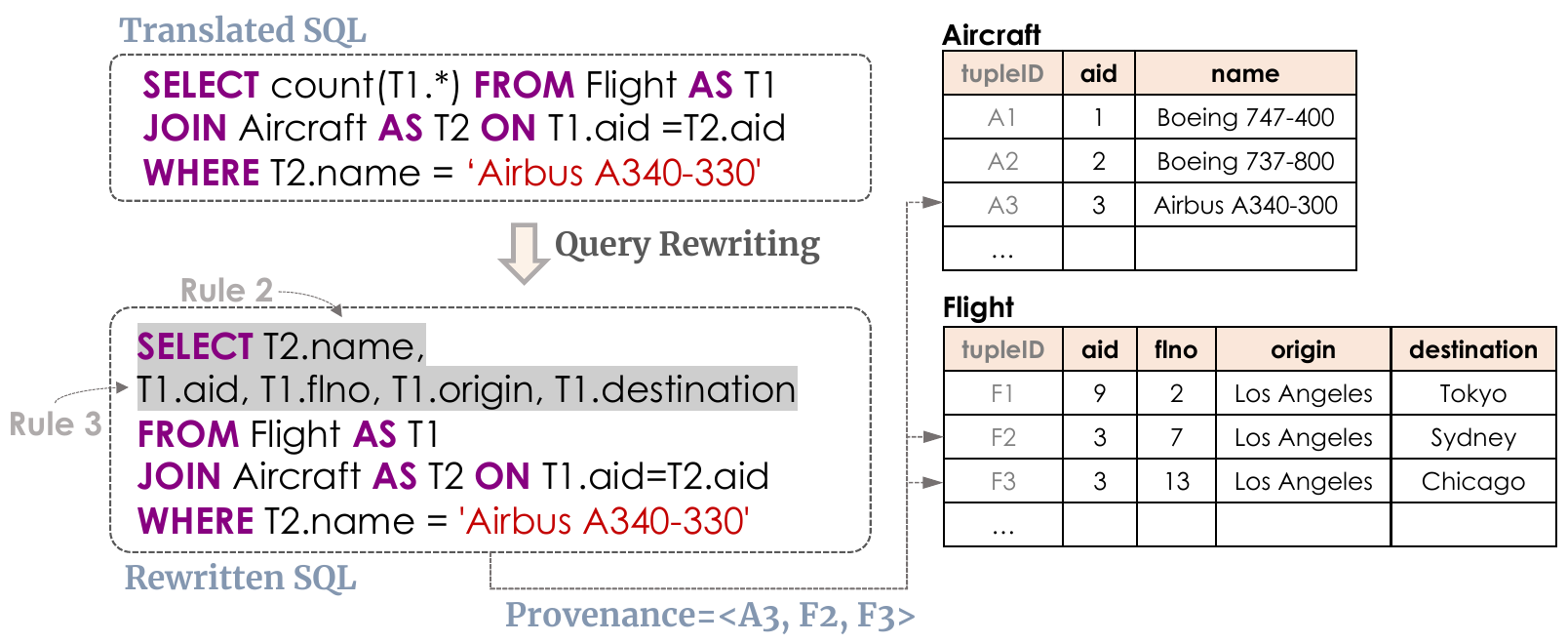}
 \vspace{-6mm}
 \captionsetup{font=small}
 \caption{{Provenance querying using a rewritten SQL query.}}
 \label{fig: rewrite}
 \vspace{-5mm}
\end{figure}

 \noindent\textbf{Example 2.} {Fig.~\ref{fig: rewrite} illustrates the query rewriting process applied to the translated SQL query presented in Fig.~\ref{fig: spider query}. The rewritten SQL query is transformed by applying Rule $2$ to retrieve the column from the condition and Rule $3$ to eliminate the aggregation by replacing it with relevant columns.}

 It is important to note that as \textsc{Cyclesql} attempts to trace the provenance information of a given query result by retrieving the relevant data records, the absence of a result from an SQL query may indicate a lack of corresponding provenance information. Therefore, \textsc{Cyclesql} skips this step for \textit{empty-result queries} and uses the operation-level semantics directly (see the subsequent section) to generate the explanation instead. Exploring more effective strategies for managing empty-result queries within the context of \textsc{Cyclesql} could pose an interesting research problem.

 \subsection{Semantics Enrichment via Annotating}\label{sec 4.2}
 The main observation arising from the provenance enrichment process is that while provenance can reveal details about the query result derived from the data, it is insufficient to provide insights into why the query result satisfies the query criteria \cite{CaJaDE21}. To give a taste of the explanation that can be produced using provenance information alone, an explanation is exemplified as follows for the example in Fig.~\ref{fig: rewrite},

 \vspace{1mm}
 \noindent{\textit{$\blacktriangleright$ Flight number 7 and flight number 13 are both associated with the Airbus A340-300 aircraft.}}
 \vspace{1mm}

 \noindent Although the explanation interprets the details about those flights with ``\textit{Airbus A340-300}'' aircraft, it does not clarify how the resulting number $2$ is derived. This lack of explicit information may impede \textsc{Cyclesql} from getting a valid explanation suitable for validating the translation afterward.

 To address the limitation above, \textsc{Cyclesql} enriches the provenance by integrating the operation-level semantics of the SQL queries (both the original and rewritten SQLs) to better reflect user query intent. The fundamental concept behind this enrichment is that different parts of the query may ``contribute'' to different parts of the provenance information (e.g., the parts of a rewritten query) or play a role in transforming the query result based on different parts of the provenance (e.g., the parts of an original query). We hope to identify semantics alignment between provenance elements and query operators, ensuring all necessary semantics are properly captured and combined.
 
 Specifically, \textsc{Cyclesql} initially treats the SQL query as a text string and divides the string into chunks that correspond to each clause in the SQL query\footnote{We observe that a subquery embodies complete semantics in a SQL query. Therefore, we consider a subquery as a whole when performing the division.}. Each query clause contains its query elements, such as the column ``name'' in the {\fontfamily{pcr}\fontsize{9}{10}\selectfont \textcolor{pp}{\textbf{WHERE}}} clause, representing its semantics. \textsc{Cyclesql} then simply navigates through the provenance (table) to locate the corresponding element mentioned in the query clauses and overlays annotations atop it as their semantics labels. If a query clause contains an asterisk element (*), we place the annotation atop the table to signify global semantics across the referenced table, rather than on an individual element. The meaning of the asterisk element is subsequently determined during the explanation generation process.

\begin{figure}[h]
 \vspace{-3mm}
 \centering
 \includegraphics[width=.9\linewidth]{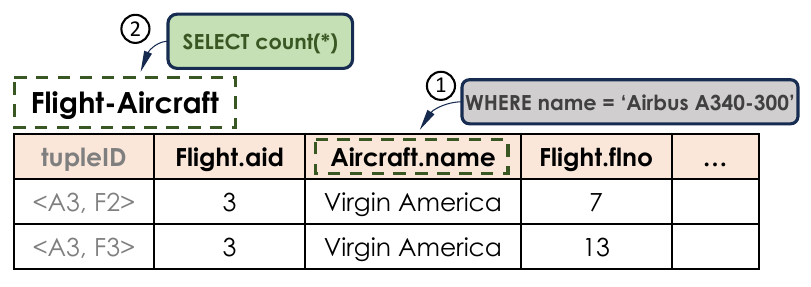}
 \vspace{-2mm}
 \captionsetup{font=small}
 \caption{{An example of the enriched provenance information.}}
 \label{fig: annotation}
 \vspace{-2mm}
\end{figure}

\noindent\textbf{Example 3.} Fig.~\ref{fig: annotation} illustrates the enriched provenance table retrieved in Fig.~\ref{fig: rewrite}, {where the table is enriched by one annotation (i.e., \circled{2}) from the original translated query, along with one annotation (i.e., \circled{1}) derived from the {\fontfamily{pcr}\fontsize{9}{10}\selectfont \textcolor{pp}{\textbf{WHERE}}} clause.}



 \subsection{Explanation Generation via Graphing}\label{sec 4.3}
 At the core of \textsc{Cyclesql} lies a basic question: how to interpret the enriched provenance information to a textual expression in free-text forms that are highly expressive and meaningful to explain the query result. Motivated by previous works on formulating provenance in NL \cite{Deutch17, Deutch20} and earlier SQL2NL works \cite{SQL2NL10, GAR23}, \textsc{Cyclesql} employs a simple but effective rule-based method to synthesize the NL explanation using the enriched provenance information.
 


 
 More specifically, \textsc{Cyclesql} initiates the process by generating a brief text summary of the query, outlining details such as the number of columns and rows in the returned result set. Next, \textsc{Cyclesql} partitions the enriched provenance table of a specific query result into distinct segments, proceeding with the incremental construction of a provenance graph. A provenance graph, denoted as $G_p (V_p, E_p)$, is a directed graph where nodes in $V_{p}$ represent provenance elements, such as the table element ``\textit{Flights}'', the column element ``\textit{filghtNo}'' and the value element ``\textit{1377}''. Each edge in $E_{p}$ is associated with a type, indicating a specific relationship between two elements. For example, as for the provenance table in Fig.~\ref{fig: annotation}, $G_{p}$ has a joint table node ``\textit{Airlines-Flights}'', which has a ``\textit{hasAttribute}'' type edge that connects with the column node ``\textit{uid}''; the column node ``\textit{uid}'', in turns, has a ``\textit{hasValue}'' type edge that connects with the value node ``\textit{12}''. Then, \textsc{Cyclesql} assigns the query annotations we obtained in Section \ref{sec 4.2} as the semantics labels to each corresponding element. Finally, \textsc{Cyclesql} traverses the graph $G_{p}$ to generate NL phrases for each provenance element with its associated labels. The NL phrases (and the initially generated summary) are then concatenated to generate the NL explanation by using some descriptive expressions (e.g., \textit{`so'}, \textit{`for'}, etc.). We present the explanation generation process in Algorithm \ref{algoritm: dialect generation}.
 

\noindent{\textbf{Example 4.} The following explanation is generated from the provenance graph in the motivating example,}

 \noindent{\textit{$\blacktriangleright$ The query returns a result with one column of aggregation type (count) and one row. For flights with aircraft, named Airbus A340-300, there are 2 flights in total.}}
 \vspace{1mm}

While the generated explanations may lack naturalness, we posit that they are informative for capturing the core semantics conveyed in the given provenance information targeted for a specific query result. These explanations can further be refined using a ``polishing model'' (e.g., a few-shot prompted LLM) to enhance their naturalness, especially for better user readability. (Refer to the evaluation in Section~\ref{5.2} for an example.)

\noindent\textbf{Join-related Semantics.} We observe that the semantics expressed in provenance may become much more abstract when derived from join-related operations \cite{GAR23}. For example, the provenance table shown in Fig.~\ref{fig: annotation} represents a new ``table'' derived from the join operation of the underlying SQL query, and the semantics of the aggregation-related semantics (i.e., {\fontfamily{pcr}\fontsize{9}{10}\selectfont\textcolor{blue}{count}} semantics) become abstract after the table joining.

\vspace{-3mm}
\SetAlCapFnt{\scriptsize}
\normalem
 \begin{algorithm}
 \SetAlgoLined
 \SetKwFunction{generateSummary}{\textsc{Generate}-SUMMARY}
 \SetKwFunction{buildGraph}{\textsc{Build}-GRAPH}
 \SetKwFunction{generatePhrase}{\textsc{Generate}-PHASE}
 \SetKwFunction{setPriority}{\textsc{Set}-PRIORITY}
 \SetKwFunction{composePhase}{\textsc{Compose}-PHASE}
  \SetKwFunction{recomposetrees}{\textsc{Recompose}-TREES}
 \SetKwInOut{KwIn}{Inputs}
 \SetKwInOut{KwOut}{Output}
 \KwIn{$R$: query result set; $T_{r}$: enriched provenance table for a query result $r$.}
 \tcp{\color{blue}\small{Summarize query statistics from the result set}}
 $s_{0}$ = \generateSummary($R$)\\
 \tcp{\color{blue}\small{Build the provenance graph for a specific result}}
 $\{p_{r_{i}}\}_{i=1}^{n} \leftarrow$ \buildGraph($T_{r}$)\\
 \For{$i = 1, \cdots, n$}{
    \tcp{\color{blue}\small{Generate NL phrase for each provenance element}}
    $s_{i} = $ \generatePhrase{$p_{r_{i}}$}\\
 }
 s = \composePhase{$s_{0}, s_{1}, \cdots, s_{n}$}\\
\KwOut{$s$: A textual string about explanation}
\caption{Natural Language Explanation Generation}
 \label{algoritm: dialect generation}
\end{algorithm}
\vspace{-3mm}

{To capture the join semantics, we introduce a heuristic method to discover the semantics before the NL phrase generation. Specifically, we represent database schemata as graph structures, with nodes as tables
and edges denoting relationships between the tables. We create an inter-table graph pool comprising pre-defined common graph topologies to capture potential semantics. For example, in a three-node graph where one node (table) connects to two other nodes (tables), this structure may represent
\textit{object-attribute} semantics.}

\begin{figure}[ht]
 \vspace{-3mm}
 \centering
 \includegraphics[width=\linewidth]{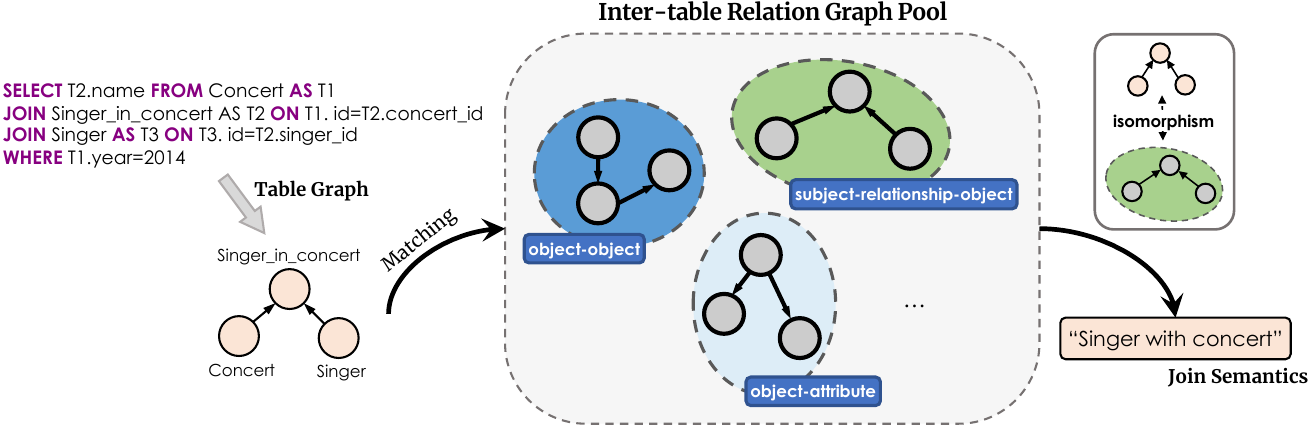}
 \vspace{-5mm}
 \captionsetup{font=small}
 \caption{Join-related semantics discovery in \textsc{Cyclesql}.}
 \label{fig: join semantics}
 \vspace{-5mm}
\end{figure}

{When a join relation appears in a query, we convert it into the graph structure as described above and then check for isomorphism with any pre-defined graphs to determine the corresponding join semantics. If no isomorphism is found, we use the associated table names to represent the join semantics instead. Fig.~\ref{fig: join semantics} shows an example of a three-table query, where the ``Singer\_in\_concert'' table has two foreign keys linked to the ``Concert'' and ``Singer'' tables. When matched with pre-defined graphs, an isomorphism graph (i.e., the three-node graph with \textit{subject-relationship-object} semantics) is identified, enabling the join semantics to be instantiated as ``\textit{singer with concert}'' using the specific table names.}



 \subsection{Explanation as NL2SQL Verifier}\label{sec 4.4}
 Once the NL explanation for a specific query result is generated, the subsequent question for \textsc{Cyclesql} is how to effectively leverage the NL explanation to validate the translation. Drawing inspiration from recent achievements in the NLP field, we thus formulate the translation validation problem as a \textit{textual entailment task} (\textit{a.k.a.} NLI task) \cite{Dagan05, SharmaSB15} and employ deep learning techniques to construct an NLI model for this purpose. Here, the NLI model aims to evaluate the semantic coherence between a given NL query and the NL explanation associated with a query result. The NL query fragment, labeled as \textit{hypothesis} ($H$), is inferred from the NL explanation fragment known as \textit{premise} ($P$)\footnote{To provide extra context, we also include the query result along with the SQL query within the premise, using a special token ($|$) to separate each part.}. In other words, an explanation $P$ entails an NL query $H$, if $H$ is considered to be true within the context of the corresponding text $P$.

 \noindent\textbf{Remark.} To implement the NLI model, popular choices include utilizing ready-to-use LLMs \cite{GPT4, LLaMA23}, off-the-shelf NLI models \cite{SemBERT20, EFL21}, or crafting a custom NLI model from scratch. However, we found that the first two choices may yield inferior performance when contrasted with the latter. (Refer to the experiments presented in Section ~\ref{5.1.2} for further details.) Therefore, in this paper, we introduce a dedicated NLI model, as outlined below, to serve as the verifier instead.

\begin{figure}[ht]
    \vspace{-3mm}
    \centering
    \includegraphics[width=\linewidth]{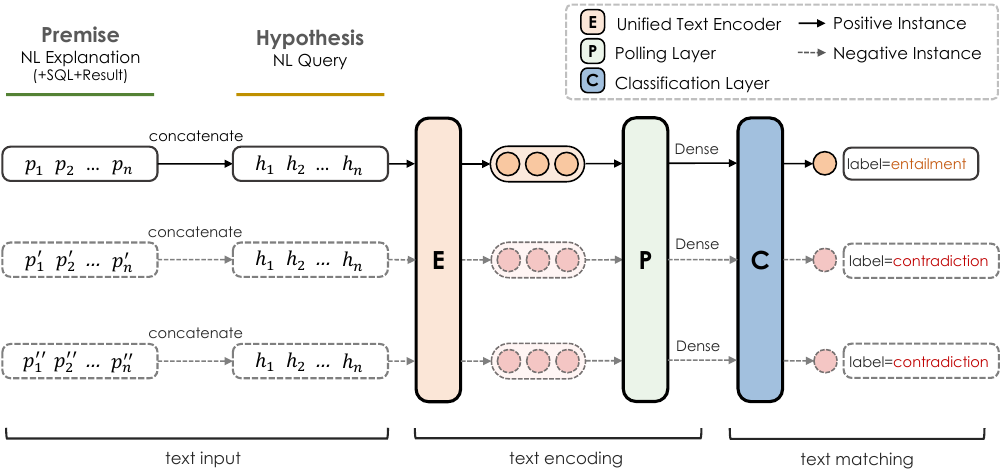}
    \vspace{-5mm}
    \captionsetup{font=small}
    \caption{NLI model with entailment and contradiction pairs.}
    \label{fig: textual entailment model}
    \vspace{-3mm}
\end{figure}

 Fig.~\ref{fig: textual entailment model} presents the architecture of the dedicated NLI model. Building upon prior works \cite{Nogueira20,rankT5}, we adopt a similar approach to construct the model, refining it via fine-tuning for specific classification objectives. We use a unified text encoder to jointly encode the hypothesis and the premise, applying a pooling layer to create a unified embedding vector, which is then fed into a classification layer for output.

\noindent\textbf{Training Data.} The training data is enforced as a set of premise-hypothesis-label triples, denoted as $\{(p_{i}, h_{i}, l_{i})\}_{i=1}^{N}$, where $p_{i}$ is an explanation (inclusive of the associated query result and SQL query), $h_{i}$ is an NL query, and $l_{i}$ is a classification label, with possible values including ``\textit{entailment}'' ($+1$) and ``\textit{contradiction}'' ($-1$). We collect the data as follows: 

\begin{itemize}[leftmargin=*]
    \item For \textit{positive samples} (i.e., instances with the  ``entailment'' label), we create samples using human-curated data (NL-SQL pairs) from public benchmarks. Initially, we execute the provided SQL query against the underlying database to get the query results. A specific query result is then randomly selected to generate the explanation using \textsc{Cyclesql}. Thus, a triple $(p_{+}, h, +1)$ is formed.
    
    \item For \textit{negative samples} (i.e., instances labeled as ``contradiction''), we generate samples from the erroneous translations produced by existing NL2SQL models on the public benchmarks. Rather than relying on human-labeled ground-truth SQL queries in public benchmarks, we initiate the process with the ``incorrect'' SQL queries generated by existing NL2SQL models and employ a similar approach for positive counterparts to complete the generation. Thus, a triple $(p_{-}, h, -1)$ is then formed.
\end{itemize}

\noindent {In this work, we use the public benchmark’s training set to build the verifier’s training set (details in the next section).}

\noindent\textbf{Loss Function.} Cross-entropy loss is a well-adopted loss function in classification tasks. However, owing to the significantly greater number of negative samples collected from various translation models compared to the positive ones from human-curated benchmarks, the label distribution in the training set becomes highly imbalanced, resulting in significant training bias. To alleviate this issue, we utilize the \textit{focal loss} \cite{focal20} as our classification loss, which is defined in the following way,

\begin{equation}
\begin{aligned}
FL(p_{t}) = -(1-p_{t})^{\gamma}\cdot\log(p_{t}) \\
p_{t} = \left\{ \begin{array}{ll}
         p & if\;\;y = 1\\ 
         1 - p & \mathrm{otherwise.}
                \end{array}\right.
\end{aligned}
\end{equation}

\noindent where $y \in {\pm 1}$ specifies the ground-truth class and $p \in [0, 1]$ is the model’s estimated probability for the class with label $y=1$, and $\gamma$ is the focusing parameter controlling the down-weighting of well-classified examples.

\section{Evaluation}\label{section 5}
This section initiates with an experimental evaluation conducted by applying \textsc{Cyclesql} to state-of-the-art NL2SQL models, thereby gauging its impact on their overall performance improvement. Following this, a qualitative evaluation incorporating a case study and a user study is carried out to evaluate the quality of explanations produced by \textsc{Cyclesql}.


\subsection{Experimental Evaluation}\label{5.1}



\subsubsection{Experimental Setup} Following are the details:

\noindent\textbf{Datasets.} We conduct experiments on the \textsc{Spider} dataset and its three robustness variants, namely \textsc{Spider-realistic}, \textsc{Spider-syn}, and \textsc{Spider-dk}, {as well as on the complex scientific \textsc{ScienceBenchmark} benchmarks.}



\begin{itemize}[leftmargin=*]
\item\textbf{\textsc{Spider}} \cite{Yu18} includes $10,181$ NL queries and $5,693$ unique SQL queries of $206$ databases covering $138$ domains. As it uses different databases across its train, dev, and test data sets, the model's generalizability can be properly evaluated.

\noindent Since its test set is not available in public, our experiments primarily focused on \textsc{Spider} validation set. We apply \textsc{Cyclesql} to \textsc{Resdsql} and \textsc{Gpt-3.5-turbo} and get the evaluation results on the test set (See Table~\ref{tab: CycleSQL overall accuracy results}).

\item\textbf{\textsc{Spider-realistic} \cite{spider-realistic21}} poses the challenges associated with text-table alignments, creating a more realistic scenario by omitting explicit column name mentions. This requires models to adeptly map NL terms to relevant schema items.

\item\textbf{\textsc{Spider-syn} \cite{spider-syn21}} modifies original NL queries of \textsc{Spider} by substituting schema-related terms with handpicked synonyms. This challenges the reliance on lexical matching.

\item\textbf{\textsc{Spider-dk} \cite{spider-dk21}} necessitates the models to know about domain-specific knowledge for the SQL generation. Preliminary observations reveal that current models encounter difficulties meeting this elevated domain-specific demand.

\item {\textbf{\textsc{ScienceBenchmark}} \cite{sciencebenchmark23} serves as a complex benchmark for three real-world scientiﬁc databases. For this benchmark, domain experts crafted $103/100/100$ high-quality NL-SQL pairs for each domain, then augmented with synthetic data generated using \textsc{Gpt-3.5}.}
\end{itemize}

\newcolumntype{?}{!{\vrule width 1pt}}
\definecolor{darkgreen}{rgb}{0, 0.70, 0}
\definecolor{darkred}{rgb}{0.70, 0, 0}
\definecolor{Gray}{gray}{0.90}
\definecolor{babyblue}{rgb}{0.54, 0.81, 0.94}
\begin{table*}[ht]
\begin{adjustbox}{width=\textwidth}
\begin{threeparttable}
  \centering
  \captionsetup{font=small, skip=-.3mm}
  \caption{{Overall translation results on \textsc{Spider} and its three variants, as well as \textsc{Sciencebenchmark} benchmarks (\%).  The fire \twemoji[height=1.0em]{fire} symbol indicates training the NLI model with the respective training data of the underlying benchmark, while the ice \twemoji[height=1.0em]{snowflake} symbol denotes freezing the model weights.}}
  \begin{tabular}{c l c c c c c c c c c c c c c c c c c} \hline
     \multicolumn{2}{c}{\multirow{3}{*}{\textbf{Models}}} & \multicolumn{5}{?c}{\cellcolor{rd!20}\twemoji[height=1.0em]{fire}\textbf{\textsc{Spider}}} &  \multicolumn{8}{?c}{\cellcolor{babyblue!30}\twemoji[height=1.0em]{snowflake}\textbf{$\textsc{Spider Variants}$}}  &
     \multicolumn{3}{?c}{\cellcolor{babyblue!30}\twemoji[height=1.0em]{snowflake}{\textbf{$\textsc{ScienceBenchmark}$}}} \\ 
     \cline{3-18}
     & & \multicolumn{3}{?c}{\textbf{\textsc{Validation}}}  & \multicolumn{2}{?c}{\textbf{\textsc{Test}}}  & \multicolumn{3}{?c}{\textbf{\textsc{Realistic}}} & \multicolumn{3}{?c}{\textbf{\textsc{Syn}}} & \multicolumn{2}{?c}{\textbf{$\textsc{Dk}$}} & \multicolumn{1}{?c}{{\textsc{OncoMx}}} & {\textsc{Cordis}} & {\textsc{Sdss}} \\
     & & \multicolumn{1}{?c}{\textbf{EM}} & \textbf{EX} & \textbf{TS} & \multicolumn{1}{?c}{\textbf{EM}} & \textbf{EX} & \multicolumn{1}{?c}{\textbf{EM}} & \textbf{EX} & \textbf{TS} & \multicolumn{1}{?c}{\textbf{EM}} & \textbf{EX} & \textbf{TS} & \multicolumn{1}{?c}{\textbf{EM}} & \textbf{EX} & \multicolumn{1}{?c}{{\textbf{EX}}} & {\textbf{EX}} & {\textbf{EX}} \\
    \hline\hline
    
    \rowcolor{Gray} \multicolumn{18}{c}{Seq2seq-based Models} \\ \hline\hline
    
    
    \multirow{2}{*}{\textsc{\textbf{SmBoP}}} & Base  & \multicolumn{1}{?c}{\textbf{74.5}} & 77.8 & 72.5 & \multicolumn{1}{?c}{69.5} & 71.1 & \multicolumn{1}{?c}{\textbf{60.2}} & 61.4 & 56.1 & \multicolumn{1}{?c}{\textbf{60.2}} & 64.9 & 58.3 & \multicolumn{1}{?c}{\cellcolor{darkgreen!18}\textbf{54.0}} & 59.1 & \multicolumn{1}{?c}{{16.2}} & {\textbf{16.0}} & {7.0} \\
    & +\textsc{Cyclesql}   & \multicolumn{1}{?c}{74.1} & \textbf{78.8} & \textbf{73.1} & \multicolumn{1}{?c}{-} & - & \multicolumn{1}{?c}{59.8} & \textbf{62.2} & \textbf{56.7} & \multicolumn{1}{?c}{59.1} & \textbf{65.5} & \textbf{58.6} & \multicolumn{1}{?c}{53.3} & \textbf{59.8} & \multicolumn{1}{?c}{{\textbf{17.2}}} & {\textbf{16.0}} & {\textbf{8.0}} \\\hline
    
    \multirow{2}{*}{$\textsc{\textbf{Picard}}_{\textsc{\textbf{3b}}}$} & Base & \multicolumn{1}{?c}{75.9} & 79.3 & \textbf{72.8} & \multicolumn{1}{?c}{72.4} & 75.1 & \multicolumn{1}{?c}{\textbf{68.9}} & 71.7 & \textbf{64.6} & \multicolumn{1}{?c}{65.6} & 69.8 & 63.2 & \multicolumn{1}{?c}{49.7} & 63.2 & \multicolumn{1}{?c}{{32.3}} & {\textbf{26.0}} & {\textbf{8.0}} \\
    & +\textsc{Cyclesql} & \multicolumn{1}{?c}{\textbf{76.1}} & \textbf{79.9} & \textbf{72.8} & \multicolumn{1}{?c}{-} & - & \multicolumn{1}{?c}{68.7} & \textbf{71.9} & \textbf{64.6} & \multicolumn{1}{?c}{\cellcolor{darkgreen!18}\textbf{65.9}} & \textbf{70.8} & \textbf{63.6} & \multicolumn{1}{?c}{\textbf{50.3}} & \textbf{63.6} & \multicolumn{1}{?c}{{\textbf{33.3}}} & {24.0} & {\textbf{8.0}} \\ \hline
    
    \multirow{2}{*}{$\textsc{\textbf{Resdsql}}_{\textsc{\textbf{Large}}}$} & Base  & \multicolumn{1}{?c}{74.0} & 77.5 & 71.6 & \multicolumn{1}{?c}{66.8} & 73.5 & \multicolumn{1}{?c}{67.3} & 69.5 & 61.6 & \multicolumn{1}{?c}{58.8} & 65.3 & 59.6 & \multicolumn{1}{?c}{50.1} & 58.1 & \multicolumn{1}{?c}{{33.0}} & {29.0} & {4.0} \\ 
    & +\textsc{Cyclesql}  & \multicolumn{1}{?c}{\textbf{75.1}} & \textbf{80.5} & \textbf{74.0} & \multicolumn{1}{?c}{\textbf{68.9}} & \textbf{77.1} & \multicolumn{1}{?c}{\textbf{69.1}} & \textbf{73.6} & \textbf{64.2} & \multicolumn{1}{?c}{\textbf{61.6}} & \textbf{69.6} & \textbf{63.2} & \multicolumn{1}{?c}{\textbf{50.5}} & \cellcolor{darkgreen!18}\textbf{61.5} & \multicolumn{1}{?c}{{\textbf{35.0}}} & {\cellcolor{darkgreen!18}\textbf{32.0}} & {\textbf{8.0}} \\ \hline

    \multirow{2}{*}{$\textsc{\textbf{Resdsql}}_{\textsc{\textbf{3b}}}$} & Base & \multicolumn{1}{?c}{76.0} & 79.4 & 73.5 & \multicolumn{1}{?c}{70.2} & 78.4 & \multicolumn{1}{?c}{68.9} & 72.2 & 64.8 & \multicolumn{1}{?c}{62.9} & 69.5 & 63.2 & \multicolumn{1}{?c}{52.0} & 59.3 & \multicolumn{1}{?c}{{34.1}} & {28.0} & {6.0} \\ 
    & +\textsc{Cyclesql}  & \multicolumn{1}{?c}{\cellcolor{darkgreen!18}\textbf{76.8}} & \cellcolor{darkgreen!18}\textbf{82.0} & \cellcolor{darkgreen!18}\textbf{76.0} & \multicolumn{1}{?c}{\cellcolor{darkgreen!18}\textbf{72.5}} & \cellcolor{darkgreen!18}\textbf{81.6} & \multicolumn{1}{?c}{\cellcolor{darkgreen!18}\textbf{71.3}} & \cellcolor{darkgreen!18}\textbf{76.5} & \cellcolor{darkgreen!18}\textbf{67.5}  & \multicolumn{1}{?c}{\textbf{65.5}} & \cellcolor{darkgreen!18}\textbf{73.3} & \cellcolor{darkgreen!18}\textbf{66.6} & \multicolumn{1}{?c}{\textbf{52.5}} & \textbf{61.3} & \multicolumn{1}{?c}{{\cellcolor{darkgreen!18}\textbf{37.4}}} & {\textbf{31.0}} & {\textbf{8.0}} \\ \hline\hline
    

    \rowcolor{Gray} \multicolumn{18}{c}{LLM-based Models} \\ \hline\hline
    
    
    \multirow{2}{*}{\makecell{\textsc{\textbf{Gpt-3.5-turbo}} \\ \small{(\textsc{5-shots})}}} & Base & \multicolumn{1}{?c}{43.8} & 72.8 & 61.7 & \multicolumn{1}{?c}{44.7} & 69.4 & \multicolumn{1}{?c}{39.4} & 65.4 & 52.8 & \multicolumn{1}{?c}{34.0} & 61.0 & 49.8 & \multicolumn{1}{?c}{39.1} & 60.6 & \multicolumn{1}{?c}{{34.3}} & {30.0} & {10.0} \\
    & +\textsc{Cyclesql} & \multicolumn{1}{?c}{\textbf{49.2}} & \textbf{77.8} & \textbf{66.2} & \multicolumn{1}{?c}{\cellcolor{darkgreen!18}\textbf{50.6}} & \cellcolor{darkgreen!18}\textbf{75.1} & \multicolumn{1}{?c}{\textbf{46.5}} & \cellcolor{darkgreen!18}\textbf{71.1} & \textbf{56.7} & \multicolumn{1}{?c}{\textbf{40.0}} & \textbf{66.7} & \textbf{55.4} & \multicolumn{1}{?c}{\textbf{43.7}} & \textbf{65.6} & \multicolumn{1}{?c}{{\textbf{43.4}}} & {\textbf{34.0}} & {\textbf{12.0}} \\\hline 

    \multirow{2}{*}{\makecell{\textsc{\textbf{Gpt-4}} \\ \small{(\textsc{5-shots})}}} & Base & \multicolumn{1}{?c}{51.9} & 77.9 & 71.2 & \multicolumn{1}{?c}{-} & - & \multicolumn{1}{?c}{46.7} & 68.6 & 55.3 & \multicolumn{1}{?c}{45.2} & 75.0 & 64.4 & \multicolumn{1}{?c}{53.7} & \textbf{68.5} & \multicolumn{1}{?c}{{51.5}} & {40.0} & {14.0} \\
    & +\textsc{Cyclesql} & \multicolumn{1}{?c}{\textbf{58.7}} & \textbf{79.8} & \textbf{73.1} & \multicolumn{1}{?c}{-} & - & \multicolumn{1}{?c}{\cellcolor{darkgreen!18}\textbf{49.0}} & \textbf{70.6} & \cellcolor{darkgreen!18}\textbf{56.9} & \multicolumn{1}{?c}{\cellcolor{darkgreen!18}\textbf{46.2}} & \cellcolor{darkgreen!18}\textbf{76.0} & \cellcolor{darkgreen!18}\textbf{66.3} & \multicolumn{1}{?c}{\cellcolor{darkgreen!18}\textbf{57.4}} & \textbf{68.5} & \multicolumn{1}{?c}{{\textbf{56.6}}} & {\textbf{43.0}} & {\textbf{15.0}} \\\hline 

    \multirow{2}{*}{{\textsc{\textbf{Chess}}}} & {Base}  & \multicolumn{1}{?c}{{23.4}} & {41.1} & {37.5} & \multicolumn{1}{?c}{{-}} & {-} & \multicolumn{1}{?c}{{21.2}} &{39.7} & {36.6} & \multicolumn{1}{?c}{{17.4}} & {37.0} & {32.5} & \multicolumn{1}{?c}{{19.5}} & {35.6} & \multicolumn{1}{?c}{{64.6}} & {46.0} & {\textbf{40.0}} \\
    & {+\textsc{Cyclesql}}   & \multicolumn{1}{?c}{{\textbf{25.6}}} & {\textbf{42.3}} & {\textbf{38.2}} & \multicolumn{1}{?c}{{-}} & {-} & \multicolumn{1}{?c}{{\textbf{23.1}}} & {\textbf{41.1}} & {\textbf{37.9}} & \multicolumn{1}{?c}{{\textbf{19.3}}} & {\textbf{38.1}} & {\textbf{33.0}} & \multicolumn{1}{?c}{{\textbf{20.2}}} & {\textbf{36.1}} & \multicolumn{1}{?c}{{\cellcolor{darkgreen!18}\textbf{65.7}}} & {\cellcolor{darkgreen!18}\textbf{52.0}} & {\cellcolor{darkgreen!18}\textbf{40.0}} \\\hline
    
    \multirow{2}{*}{$\textsc{\textbf{Dailsql}}_{\textsc{\textbf{3.5}}}$} & Base & \multicolumn{1}{?c}{65.1} & 81.3 & 74.1 & \multicolumn{13}{?c}{\cellcolor{gray!20}}  \\
    & +\textsc{Cyclesql} & \multicolumn{1}{?c}{\cellcolor{darkgreen!18}\textbf{67.2}} & \cellcolor{darkgreen!18}\textbf{81.8} & \cellcolor{darkgreen!18}\textbf{74.3} & \multicolumn{13}{?c}{\cellcolor{gray!20}} \\\hline    
  \end{tabular}
\label{tab: CycleSQL overall accuracy results}
\end{threeparttable}
\end{adjustbox}
\vspace{-5mm}
\end{table*}

\noindent\textbf{Baseline NL2SQL Models.} We consider the following two types of baselines: \textit{Seq2seq-based translation models} fine-tuned on the \textsc{Spider} training set, including \textsc{SmBoP} \cite{smbop21}, \textsc{Picard} \cite{PICARD21}, and \textsc{Resdsql} \cite{resdsql23}, and \textit{LLM-based models}, such as \textsc{Gpt-3.5-turbo}, \textsc{Gpt-4}, {\textsc{Chess} \cite{chess24}}, and \textsc{Dailsql} \cite{DAILSQL23}. We briefly describe each baseline model as follows.

\begin{itemize}[leftmargin=*]
    \item \textbf{\textsc{SmBoP}} {re-formats the top-down decoding method and employs a reverse bottom-up decoding mechanism based on \textsc{Grappa} \cite{Grappa21}, a pre-trained model for semantic parsing.}
    

    \item \textbf{\textsc{Picard}} constrains the auto-regressive decoders via incremental parsing. We denote as $\textsc{Picard}_{\textsc{3b}}$ as it is implemented based on \textsc{T5-3b} model \cite{T5}.

    
    \item \textbf{\textsc{Resdsql}} encodes the question and schema as a tagged sequence and is implemented with T5 model in \textit{Base}, \textit{Large} and \textit{3B} scales. Our experiments use \textsc{Resdsql} with the latter two scales, denoted to as $\textsc{Resdsql}_{\textsc{Large}}$ and $\textsc{Resdsql}_{\textsc{3b}}$, and exclude its combination with \textsc{Natsql} \cite{Natsql21}.
    
    \item {\textbf{\textsc{Gpt-3.5-turbo}}}\footnote{{Specifically refers to the \texttt{\textsc{Gpt-3.5-turbo-0125}} version of the model.}} and \textbf{\textsc{Gpt-4}} stand out as two of the most advanced LLMs. {In our experiment, we perform experiments on the two models using \textit{5-shot prompting} with a simple instruction as depicted in the below example,}

    \begin{figure}[ht]
      \vspace{-4mm}
      \begin{tikzpicture}
      \node (table) [inner sep=0pt] {
      \begin{adjustbox}{width=\linewidth}
        \begin{tabular}{l} 
         \\ [-2.0ex]
         \rowcolor{gray!10}\makecell[l]{\textbf{\#\#\#Instruction} \\
         Given a database schema and NL question, and generate an SQL query. \\
         \textbf{\#\#\#Schema} \\
         Table Player with columns 'pID', 'pName', 'yCard', 'HS'; \\
         Table Tryout with columns 'pID', 'cName', 'pPos', 'decision';\\
         \textbf{\#\#\#NL Question} \\
         For each position, what is the maximum number of hours for students \\
         who spent more than 1000 hours training?\\
         \textbf{\#\#\#Demonstrations} \\
         (demonstrations)\\\\
         \textbf{\#\#\#Target SQL Query}} \\
      \end{tabular}
      \end{adjustbox}
      };
      \draw[line width=0.30mm, rounded corners=.1em](table.north west) rectangle (table.south east);
      \end{tikzpicture}
      \vspace{-7mm}
      \label{fig: prompt}
    \end{figure}

    \item {\textbf{\textsc{Chess}} introduces a pipeline that retrieves relevant data, selects an efficient schema, and synthesizes correct SQLs. We use its published fine-tuned model\footnote{{\url{https://huggingface.co/AI4DS/NL2SQL_DeepSeek_33B}}} for candidate generation and \textsc{Gpt-3.5-turbo} for the remaining LLM calls.}
    
    \item \textbf{\textsc{Dailsql}} utilizes demonstration selection by analyzing NL and SQL query similarity. In our experiments\footnote{We fail to reproduce its preliminary model, \textsc{Graphix} \cite{Graphix23}, for pre-generating queries. Therefore, our evaluation only uses the publicly available pre-generated queries from the authors on the \textsc{Spider} validation set.}, we use \textsc{Dailsql} with \textsc{Gpt-3.5-turbo}, denoted as $\textsc{Dailsql}_{\textsc{3.5}}$.
    
\end{itemize}

\noindent Note that all three Seq2seq-based models are explicitly designed to handle specific values in SQL queries. Thus, they are capable of generating valid (executable) SQL queries, allowing effective provenance tracking by \textsc{Cyclesql}.

\noindent\textbf{Training Settings.} The NLI model is initialized by the \textsc{T5-large} model, incorporating a sequence classification/head on top, achieved through a linear layer over the pooled output. We modify the default cross-entropy loss used in the T5 model to the focal loss, setting the focusing parameter $\gamma$ and the weighted factor $\alpha$ to $2.0$ and $0.75$, respectively. Additionally, we re-scale the classification weights of the two classes to $2.7$ and $1.0$, respectively, accounting for the imbalanced training data distribution. We then utilize the Adam optimizer with a learning rate of $5e-6$ to train the model.

\noindent \textit{Training data.} {We use the baseline models to generate negative samples by extracting erroneous translations from their outputs on the \textsc{Spider} training data. Here, a translation is considered erroneous if the execution result of the translated SQL diverges from the ground truth. We used the script provided by \textsc{Spider} authors to determine the equivalence of two different tuple sets based on ``bag semantics'', making the order irrelevant. As a result, we collected about $30,000$ training queries, including positive samples derived from the \textsc{Spider} training set.}

 \noindent\textbf{Inference Settings.} To establish the feedback loop in the end-to-end translation process, we apply \textsc{Cyclesql} to the baseline models in the following manner: Initially, each baseline model generates a list of top-$k$ candidate SQL queries for a given NL query\footnote{For those Seq2seq-based models, we use beam search and set $k$ to $8$; For those LLM-based counterparts, we specify the API parameter to $5$ instead.}. Subsequently, \textsc{Cyclesql} is applied to iteratively validate each candidate SQL query of a baseline model. If the validation succeeds, the candidate SQL query is considered as the final translation result for the given NL query. Otherwise, \textsc{Cyclesql} repeats the process with the next ranked candidate until no more candidate SQL queries remain to check. Notably, if all candidate SQL queries from a baseline model fail validation for the given NL query, the top-$1$ candidate SQL query is designated as the outcome.

\noindent\textbf{Evaluation Metrics.}
The metrics are \textit{syntactic accuracy} (EM), \textit{execution accuracy} (EX), and \textit{test suite accuracy} (TS). 

\begin{itemize}[leftmargin=*]
    \item \textbf{Syntactic Accuracy (EM)} measures whether the generated query exactly matches the ground truth, ignoring specific values in the SQL statements.

    
    \item \textbf{Execution Accuracy (EX)} evaluates if the result matches the ground truth by executing the generated SQL query. 
    
    
    \item \textbf{Test Suite Accuracy (TS)} refines the \textit{EX metric} through a distilled test suite of databases, examining if the translated query passes all EX assessments across these distilled databases. For evaluation, we utilize the script provided by \cite{ruiqi20} to generate an augmented $100$-fold distilled database.
\end{itemize}

\subsubsection{Results on \textsc{Spider}} Table~\ref{tab: CycleSQL overall accuracy results} presents our main results.

 \noindent\textsc{\textbf{Cyclesql}} \textbf{consistently improves over all base models}, particularly evident in the execution and test suite accuracy metrics\footnote{Given that the NLI model considers a translation as correct based on the execution equivalence criteria (as described in Training Settings in this section), we may expect some fluctuations in the syntactic accuracy evaluation.}. One remarkable outcome is observed when applying to $\textsc{Resdsql}_{\textsc{3b}}$, obtaining an execution accuracy of 82.0\% on the validation set and 81.6\% on the test set of \textsc{Spider}. {\textit{This performance surpasses the best-reported result among leading Seq2seq transformer-based models on the leaderboard.}}

\definecolor{Gray}{gray}{0.85}

It is worth noting that, unlike Seq2seq-based models, \textsc{Cyclesql} significantly enhances the syntactic accuracies of LLM-based counterparts. This notable difference stems from the fact that LLMs serve as NL2SQL models without tailored fine-tuning on existing benchmarks. This inherent diversity in the generation can be complemented by \textsc{Cyclesql} (i.e., training on benchmark-specific data), enabling LLMs to align more effectively with benchmark-specific targeted outputs.

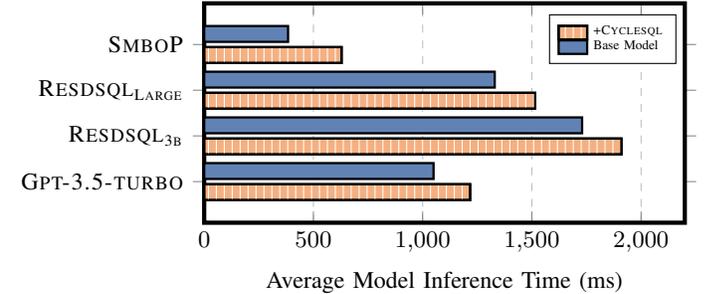
\begin{figure}[ht]
    \vspace{-5mm}
    \begin{subfigure}[b]{1.0\linewidth}
    \centering
    \captionsetup{font=small}
    \caption{Average iterations of \textsc{Cyclesql} on \textsc{Spider} validation set}
    \label{tab: average iterations}
    \begin{adjustbox}{width=\linewidth}
     \begin{tabular}{|c|c|c|c|c|c|} \hline
    \multirow[b]{2}{*}{\textbf{\makecell{Average \\ Iterations}}} & \makecell{\textsc{Smbop}\\ (+\textsc{Cyclesql})} & \makecell{\textsc{Picard}\\ (+\textsc{Cyclesql})} &
    \makecell{$\textsc{Resdsql}_{\textsc{Large}}$\\ (+\textsc{Cyclesql})} & \makecell{$\textsc{Resdsql}_{\textsc{3b}}$ \\(+\textsc{Cyclesql})} & \makecell{\textsc{Gpt-3.5-turbo} \\ (+\textsc{Cyclesql})} \\ 
    \cline{2-6}
      & 2.44 & 3.78 & 1.90 & 1.88 & 1.87 \\
    \hline
     \end{tabular}
    \end{adjustbox}
    \end{subfigure}
    \par\bigskip
    \begin{subfigure}[b]{.9\linewidth}
    \centering    
    \begin{tikzpicture}
    \definecolor{tropicalrainforest}{rgb}{0.66, 0.82, 0.56}
    \definecolor{trolleygrey}{rgb}{0.5, 0.5, 0.49}
   	\definecolor{iryellow}{rgb}{1.0, 0.84, 0.38}
    \definecolor{glaucous}{rgb}{0.38, 0.51, 0.71}
    \definecolor{orange}{rgb}{0.96, 0.69, 0.51}
    \definecolor{violet(ryb)}{rgb}{0.53, 0.39, 0.69}
    \begin{axis}[
        height=4.5cm,
        width=\linewidth,
        xbar = .70mm,
        xmin = 0,
        xmax = 2200,
        axis line style=ultra thick,
        enlarge y limits=0.3,
        xlabel={Average Model Inference Time (ms)},
        xlabel style={font=\small},
        symbolic y coords={\textsc{Gpt-3.5-turbo},$\textsc{Resdsql}_{\textsc{3b}}$,$\textsc{Resdsql}_{\textsc{Large}}$,\textsc{SmboP}},
        xticklabel style={font=\small},
        yticklabel style={font=\small},
        ytick=data,
    	bar width=6pt,
    	xmajorgrids=true,
    	grid style=dashed,
        nodes near coords align={vertical},
        legend image code/.code={%
            \draw[#1, draw=black] (0cm,-0.1cm) rectangle (0.38cm,0.08cm);
        },
        legend cell align={left},
        legend style={
                font=\tiny, 
                at={(0.85,0.95)},
                anchor=north,
                row sep=-0.12cm,
            }
        ]
        \addplot [
            line width= .3mm,
            fill=orange,
            postaction={
                pattern=vertical lines, 
                pattern color=white
            }
        ]
        coordinates {(629,\textsc{SmboP}) (1515,$\textsc{Resdsql}_{\textsc{Large}}$) (1910,$\textsc{Resdsql}_{\textsc{3b}}$) (1217.5,\textsc{Gpt-3.5-turbo})};
        \addplot [
            line width=.3mm,
            fill=glaucous,
        ]
        coordinates {(384,\textsc{SmboP}) (1330,$\textsc{Resdsql}_{\textsc{Large}}$) (1730,$\textsc{Resdsql}_{\textsc{3b}}$) (1050,\textsc{Gpt-3.5-turbo})};
        \legend{+\textsc{Cyclesql}, Base Model}
    \end{axis}
    \end{tikzpicture}
    \vspace{-7mm}
    \captionsetup{font=small}
    \caption{Inference latency comparison with or without \textsc{Cyclesql}}
    \label{fig: latency results}
    \end{subfigure}
\vspace{-1mm}
\captionsetup{font=small}
\caption{Scalability evaluation on \textsc{Cyclesql}.}
\vspace{-2mm}
\end{figure}

On the other hand, applying \textsc{Cyclesql} to \textsc{Picard} yields limited performance gains, mainly due to lower-quality sampled queries. That is, \textsc{Picard} struggles to generate the correct SQL within the top sampled outputs (See Table~\ref{tab: average iterations} for more details), requiring multiple attempts to find the correct query. This emphasizes the need for the NLI model used in \textsc{Cyclesql} to robustly identify preceding incorrect translations. {In addition, \textsc{Chess} exhibits poorer performance compared to other baselines. Upon examining the failed cases, we found that this is due to limitations in the evaluation script with semantically equivalent queries: \textsc{Chess} tends to generate ``ID-like'' projection columns, which the evaluation marks as incorrect, even when the semantics match the ground-truth (e.g., discrepancies between {\fontfamily{qcr}\fontsize{9}{10}\selectfont\textcolor{blue}{count}}(id) and {\fontfamily{qcr}\fontsize{9}{10}\selectfont\textcolor{blue}{count}}(*)).}

\noindent \textbf{Break-down Results}. Table \ref{tab: spider results by difficulty} provides a breakdown of the execution accuracy on \textsc{Spider}, categorized by the SQL difficulty levels defined in \textsc{Spider}. Not surprisingly, the performance of all the models drops with increasing difficulty. Notably, the application of \textsc{Cyclesql} leads to a consistent improvement, particularly in ``Medium'' and ``Hard'' queries for all translation models, except for a slight dip observed in ``Hard'' queries with \textsc{SmBoP}. In addition, a significant improvement is observed in ``Extra Hard'' queries across the four LLM-based models upon the application of \textsc{Cyclesql}, which, in particular, achieved 59.6\% accuracy on \textsc{Gpt-4}. This suggests that LLM-based models can generate complex queries through iterative attempts, and \textsc{Cyclesql} effectively aids in discerning and selecting queries from these successive attempts.

\begin{table}[ht]
  \vspace{-1mm}
  \centering
  \captionsetup{font=small}
  \caption{Execution accuracy (\%) by SQL difficulty levels.}
  \vspace{-2mm}
  \begin{adjustbox}{width=\linewidth}
  \begin{tabular}{c l c c c c} \hline
    \multicolumn{2}{c}{Model}  & \multicolumn{1}{?c}{\textbf{Easy}}  & \textbf{Medium} & \textbf{Hard}  & \textbf{Extra Hard} \\ \hline
    \multirow{2}{*}{\textsc{\textbf{SmBoP}}} & Base & \multicolumn{1}{?c}{90.7} & 82.7 & 70.7 & 52.4 \\

    & +\textsc{Cyclesql} & \multicolumn{1}{?c}{90.7} & $\textbf{84.1}_{(\uparrow 1.4)}$  & $69.5_{(\downarrow 1.2)}$ & $53.0_{(\uparrow 0.6)}$ \\\hline
    
    \multirow{2}{*}{$\textsc{\textbf{Picard}}_{\textsc{\textbf{3b}}}$} & Base & \multicolumn{1}{?c}{95.6} & 85.4  & 67.8 & 50.6\\
    & +\textsc{Cyclesql} & \multicolumn{1}{?c}{95.6} & $86.1_{(\uparrow 0.7)}$  & $69.5_{(\uparrow 1.7)}$ & 50.6 \\\hline

    \multirow{2}{*}{$\textsc{\textbf{Resdsql}}_{\textsc{\textbf{Large}}}$} & Base & \multicolumn{1}{?c}{92.3} & 83.4 & 66.1 & 51.2 \\
    & +\textsc{Cyclesql} & \multicolumn{1}{?c}{$93.5_{(\uparrow 1.2)}$} & $86.1_{(\uparrow 0.7)}$  & $\textbf{73.0}_{(\uparrow \textbf{6.9})}$ & $53.6_{(\uparrow 2.4)}$ \\ \hline
    
    \multirow{2}{*}{$\textsc{\textbf{Resdsql}}_{\textsc{\textbf{3b}}}$} & Base & \multicolumn{1}{?c}{94.0} & 85.7 & 65.5 & 55.4 \\
    & +\textsc{Cyclesql} & \multicolumn{1}{?c}{94.0} & $\textbf{89.0}_{(\uparrow \textbf{3.3})}$  & $\textbf{74.7}_{(\uparrow \textbf{9.2})}$ & $53.0_{(\downarrow 0.4)}$ \\ \Xhline{3\arrayrulewidth}
    
    \multirow{2}{*}{\makecell{\textsc{\textbf{Gpt-3.5-turbo}} \\ \small{(\textsc{5-shots})}}} & Base & \multicolumn{1}{?c}{84.3} & 78.5  & 65.5 & 48.2 \\
    & +\textsc{Cyclesql} & \multicolumn{1}{?c}{$86.3_{(\uparrow 2.0)}$} & $\textbf{83.0}_{(\uparrow \textbf{4.5})}$  & $\textbf{73.0}_{(\uparrow \textbf{7.5})}$ & $\textbf{56.0}_{(\uparrow \textbf{7.8})}$  \\\hline

    \multirow{2}{*}{\makecell{\textsc{\textbf{Gpt-4}} \\ \small{(\textsc{5-shots})}}} & Base & \multicolumn{1}{?c}{90.3} & 84.3  & 63.8 & 56.6 \\
    & +\textsc{Cyclesql} & \multicolumn{1}{?c}{$90.7_{(\uparrow 0.4)}$} & $85.4_{(\uparrow 1.1)}$  & $\textbf{66.7}_{(\uparrow \textbf{2.9})}$ & $\textbf{59.6}_{(\uparrow \textbf{3.0})}$ \\\hline

    \multirow{2}{*}{{\textsc{\textbf{Chess}}}} & {Base} & \multicolumn{1}{?c}{{70.2}} & {25.3} & {39.7} & {19.3} \\

    & {+\textsc{Cyclesql}} & \multicolumn{1}{?c}{{$\textbf{71.6}_{(\uparrow \textbf{1.4})}$}} & {$25.6_{(\uparrow \textbf{0.3})}$}  & {$\textbf{41.1}_{(\uparrow \textbf{1.4})}$} & {$\textbf{21.6}_{(\uparrow \textbf{2.3})}$} \\\hline
    
    \multirow{2}{*}{$\textsc{\textbf{Dailsql}}_{\textsc{\textbf{3.5}}}$} & Base & \multicolumn{1}{?c}{91.1} & 86.5  & 77.0 & 57.2 \\
    & +\textsc{Cyclesql} & \multicolumn{1}{?c}{91.1} & $\textbf{86.8}_{(\uparrow \textbf{0.3})}$  & $\textbf{77.6}_{(\uparrow \textbf{0.6})}$ & $\textbf{59.0}_{(\uparrow \textbf{1.8})}$ \\\hline
  \end{tabular}
  \end{adjustbox}
  \label{tab: spider results by difficulty}
  \vspace{-2mm}
\end{table}

\noindent\textbf{Scalability Evaluation.} To improve models' accuracies, the iterative process involved in \textsc{Cyclesql} may take extra time during the model inference. To assess its accuracy-latency tradeoff, we report the average iterations required by \textsc{Cyclesql} and compare the inference time of baseline models with or without \textsc{Cyclesql}. Note that we make the comparison in an online setting. That is, we assume that all the trained neural network models have already been loaded into the memory.

Table~\ref{tab: average iterations} shows that upon integration with \textsc{Cyclesql}, most baselines can find the ``best'' translation results within $1\sim 2$ iterations, except \textsc{Picard}, which may require about $4$ iterations due to relatively lower quality of its sampled SQL queries as we mentioned earlier. On the other hand, as depicted in Fig.~\ref{fig: latency results}, given that each baseline model can generate a list of top-K query candidates in a single inference process, the additional execution time overhead incurred by \textsc{Cyclesql} is minimal\footnote{Due to the decoding process of \textsc{Picard} necessitating interaction with a web service to validate each predicted token, the inference time of \textsc{Picard} is time-consuming. Therefore, we omit it from this comparison.}. This showcases that \textsc{Cyclesql} strikes a favorable balance between translation and inference latency in its current settings. In addition, the inference time of different models heavily depends on their respective model sizes. For example, \textsc{SmboP} with approximately $360$ million parameters, demonstrates notably faster performance compared to $\textsc{Resdsql}_{\textsc{3b}}$, which boasts around $3$ billion parameters.

\subsubsection{Results on Robustness Settings.} {To validate the robustness, we use the \textit{frozen verifier} (trained only once on the \textsc{Spider} training set) and evaluate the performance of \textsc{Cyclesql} on the three \textsc{Spider} variants and \textsc{ScienceBenchmark}. Table~\ref{tab: CycleSQL overall accuracy results} reports the comprehensive results.}

The results show that \textsc{Cyclesql} can still consistently enhance the performance of the models across all benchmarks. We attribute this to the incorporation of rich semantics from the data into the explanations used to train the verifier and thus exhibit its robustness in terms of question perturbations. {Additionally, due to the complexity of queries in \textsc{ScienceBenchmark}, most models perform poorly, with \textsc{Chess} being a notable exception, underscoring the challenge of handling queries in real-world databases.}



\subsubsection{Analysis of \textsc{Cyclesql}}\label{5.1.4} The two primary components of \textsc{Cyclesql} are \textit{feedback}, and \textit{verifier}. To gain a deeper understanding, we perform additional experiments in this section to analyze the importance of these two components. Our exploration centers around the following two key questions:
\begin{enumerate}[leftmargin=*, label={(\roman*)}]
  \item \textit{What impact does utilizing a more natural (simpler) approach for generating feedback have on \textsc{Cyclesql}?}
  \item \textit{Is the reconstructed NLI model the principled verifier for \textsc{Cyclesql} compared to other alternatives?}
\end{enumerate}

\noindent\textbf{Impact of Feedback Quality.} Self-provided feedback plays a crucial role in \textsc{Cyclesql}. To quantify its impact, we conduct a comparative experiment by adopting a simpler method (i.e., SQL2NL) to generate the feedback for \textsc{Cyclesql}: providing the generated SQL query along with the database schema and asking an LLM\footnote{In our experiment, we choose to use \textsc{Gpt-3.5-turbo} for the evaluation.} to generate the NL explanation (i.e., feedback) directly. The feedback is then used in a similar way to train the verifier under the same training settings as those employed previously. We compare the performance of \textsc{Cyclesql} with verifiers resulting from different feedback settings, by applying to $\textsc{Resdsql}_{\textsc{Large}}$ and \textsc{Gpt-3.5-turbo}.


\definecolor{tropicalrainforest}{rgb}{0.66, 0.82, 0.56}
\definecolor{trolleygrey}{rgb}{0.5, 0.5, 0.49}
\definecolor{iryellow}{rgb}{1.0, 0.91, 0.64}
\definecolor{glaucous}{rgb}{0.38, 0.51, 0.71}
\definecolor{cardinal}{rgb}{0.77, 0.33, 0.3}
\definecolor{orange}{rgb}{0.96, 0.69, 0.51}
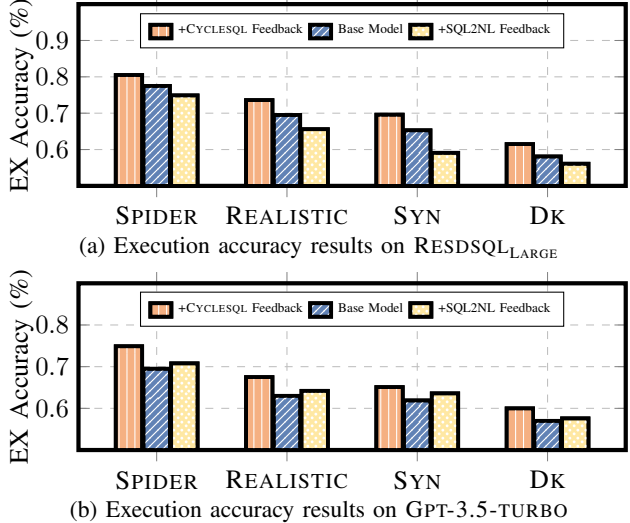
\begin{figure}[ht]
\vspace{-5mm}
\begin{subfigure}[b]{1.0\linewidth}
    \centering
    \begin{tikzpicture}
    \begin{axis}[
        height=4.0cm,
        width=\linewidth,
        ybar = .17mm, 
        enlarge y limits=0,
        enlarge x limits=0.20, 
        ymin = 0.5,
        ymax = 1.0,
        ytick={0.6,0.7,0.8,0.9},
        axis line style=ultra thick,
        xticklabel={font=\small},
        ylabel={EX Accuracy (\%)},
        ylabel style={yshift=-3ex},
        symbolic x coords={\textsc{Spider}, \textsc{Realistic}, \textsc{Syn}, \textsc{Dk}}, 
        xtick=data,
        legend image code/.code={%
            \draw[#1, draw=black] (0cm,-0.1cm) rectangle (0.28cm,0.08cm);
        },
    	bar width=10pt,
        xmajorgrids=true,
    	ymajorgrids=true,
    	grid style=dashed,
        nodes near coords align={vertical}, 
        legend pos=north east,
        legend style={
            font=\tiny,
            at={(0.5,0.95)},
            legend columns=-1,
    	    anchor=north
            }
    	]
     \addplot [
        line width= .5mm, 
        fill=orange,
        postaction={
            pattern=vertical lines, pattern color=white
        }
    ]
    coordinates {(\textsc{Spider},0.805) (\textsc{Realistic},0.736) (\textsc{Syn},0.696) (\textsc{Dk},0.615)};
    \addplot[
        line width= .5mm, 
        fill=glaucous,
        postaction={
            pattern=north east lines, pattern color=white
        }
    ]
    coordinates {(\textsc{Spider},0.775) (\textsc{Realistic},0.695) (\textsc{Syn},0.653) (\textsc{Dk},0.581)};
        
    \addplot [
        line width= .5mm,
        fill=iryellow,
        postaction={
            pattern=crosshatch dots, pattern color=white
         }
    ]
    coordinates {(\textsc{Spider}, 0.749) (\textsc{Realistic},0.656) (\textsc{Syn},0.591) (\textsc{Dk},0.561)};
    \legend{+\textsc{Cyclesql} Feedback, Base Model, +SQL2NL Feedback}
    \end{axis}
    \end{tikzpicture}
    \vspace{-2mm}
    \captionsetup{font=small}
    \caption{Execution accuracy results on $\textsc{Resdsql}_{\textsc{Large}}$}
\end{subfigure}
\bigskip
\begin{subfigure}[b]{1.0\linewidth}
    \centering
    \begin{tikzpicture}
    \begin{axis}[
        height=3.8cm,
        width=\linewidth,
        ybar = .17mm, 
        enlarge y limits=0,
        enlarge x limits=0.20, 
        ymin = 0.5,
        ymax = 0.9,
        ytick={0.6,0.7,0.8},
        axis line style=ultra thick,
        xticklabel={font=\small},
        ylabel={EX Accuracy (\%)},
        ylabel style={yshift=-3ex},
        symbolic x coords={\textsc{Spider}, \textsc{Realistic}, \textsc{Syn}, \textsc{Dk}}, 
        xtick=data,
        legend image code/.code={%
            \draw[#1, draw=black] (0cm,-0.1cm) rectangle (0.28cm,0.08cm);
        },
    	bar width=10pt,
        xmajorgrids=true,
    	ymajorgrids=true,
    	grid style=dashed,
        nodes near coords align={vertical}, 
        legend pos=north east,
        legend style={
            font=\tiny,
            at={(0.5,0.95)},
            legend columns=-1,
    	    anchor=north
            }
    	]
     \addplot [
                line width= .5mm, 
                fill=orange,
                postaction={
                    pattern=vertical lines, pattern color=white
                }
            ]
        	coordinates {(\textsc{Spider},0.749) (\textsc{Realistic},0.675) (\textsc{Syn},0.651) (\textsc{Dk},0.600)};
        \addplot[
                line width= .5mm, 
                fill=glaucous,
                postaction={
                    pattern=north east lines, pattern color=white
                }
            ]
        	coordinates {(\textsc{Spider},0.695) (\textsc{Realistic},0.630) (\textsc{Syn},0.619) (\textsc{Dk},0.570)};
        
        \addplot [
                line width= .5mm, 
                fill=iryellow,
                postaction={
                    pattern=crosshatch dots, pattern color=white
                 }
            ]
        	coordinates {(\textsc{Spider}, 0.708) (\textsc{Realistic},0.642) (\textsc{Syn},0.636) (\textsc{Dk},0.576)};
        \legend{+\textsc{Cyclesql} Feedback, Base Model, +SQL2NL Feedback}
    \end{axis}
    \end{tikzpicture}
     \vspace{-2mm}
    \caption{Execution accuracy results on \textsc{Gpt-3.5-turbo}}
\end{subfigure}
\vspace{-10mm}
\captionsetup{font=small}
\caption{Execution results on two models compared with SQL2NL.}
\label{fig: sql2nl ex results}
\vspace{-2mm}
\end{figure}

Fig.~\ref{fig: sql2nl ex results} presents the EX results across the four \textsc{Spider}-related benchmarks. It is evident that relying on the feedback generated by the SQL2NL approach becomes unreliable for the models, which may even lead to a negative impact on their performance when applying to $\textsc{Resdsql}_{\textsc{Large}}$. The findings emphasize the importance of feedback quality, demonstrating that the feedback generated by the designed approach used in \textsc{Cyclesql} embeds rich semantics, thus establishing a reliable feedback loop inside the NL2SQL process.

\noindent\textbf{Impact of Verifier Selection.}\label{5.1.2} Concerning the verifier construction crafted within \textsc{Cyclesql} (as outlined in Section \ref{sec 4.4}), a natural question may arise about the necessity of the selected design. Therefore, we select \textit{two ``strawman'' verifiers} to replace the one used in \textsc{Cyclesql}: one is a 5-shot prompted \textsc{Gpt-3.5-turbo}, and another is an off-the-shelf pre-built NLI model, named \textsc{Sembert} \cite{SemBERT20}. Table~\ref{tab: verifier comparison results} shows the translation results of \textsc{Cyclesql} embedded with different verifiers by applying to $\textsc{Resdsql}_{\textsc{3b}}$ on \textsc{Spider} validation set.

As can be seen, the performance under the current setting showcases notable enhancements compared to the other two ``strawman'' verifiers. It is noteworthy that utilizing the few-shot prompted \textsc{Gpt-3.5-turbo} can also serve as a capable verifier straight out of the box, thereby bringing some benefits to the base model. In addition, due to the disparity between NL-explanation pairs and sentence pairs in public datasets, the pre-trained \textsc{Sembert} model struggles to provide reliable verification outcomes, resulting in adverse effects.

Additionally, we compute three Oracle scores to estimate future headroom. Assuming a perfect verifier is used, which assigns the label ``entailment'' only when the translated SQL matches the ground truth, it boosts accuracy to 84.4\%, 86.9\%, and 82.5\% for the three evaluation metrics, respectively.

\begin{table}[ht]
  \vspace{-2mm}
  \centering
  \captionsetup{font=small}
  \caption{Translation results of different verifier selections.}
  \vspace{-2mm}
  \begin{adjustbox}{width=\linewidth}
  {\begin{tabular}{l c c c} \hline
    Model Variant  & \textbf{EM} & \textbf{EX} & \textbf{TS}  \\ \hline
    \textbf{Base Model ($\textsc{Resdsql}_{\textsc{3b}}$)}  & 76.0 & 79.4 & 73.5\\
    \;\; +\textsc{Cyclesql}  & $\textbf{76.8}_{(\uparrow \textbf{0.8})}$ & $\textbf{82.0}_{(\uparrow\uparrow \textbf{2.6})}$ & $\textbf{76.0}_{(\uparrow\uparrow \textbf{2.5})}$ \\
    \;\; +\textsc{Cyclesql} (w/ LLM verifier) & $74.9_{(\downarrow 1.1)}$ & $80.1_{(\uparrow 0.7)}$ & $73.8_{(\uparrow 0.3)}$\\
    \;\; +\textsc{Cyclesql} (w/ pre-built NLI verifier) & $73.0_{(\downarrow 1.9)}$  & $77.6_{(\downarrow 1.8)}$ & $71.7_{(\downarrow 1.8)}$ \\ \Xhline{3\arrayrulewidth}
    \;\; +\textsc{Cyclesql} (w/ oracle verifier) & 84.4 & 86.9 & 82.5   \\ 
    
    \hline
  \end{tabular}}
  \end{adjustbox}
  \label{tab: verifier comparison results}
  \vspace{-3mm}
\end{table}

\subsubsection{{Error Analysis}}\label{5.1.5} {To better understand \textsc{Cyclesql}, we examined the failure translations of \textsc{Cyclesql} (with $\textsc{Resdsql}_{\textsc{3b}}$) on \textsc{Spider} validation set. We identify the following two major categories for the failures.}

\begin{itemize}[leftmargin=*]
    \item {\textbf{Semantics Drift Problem.} Some translation errors stem from inaccurate verification, especially when there is a semantic divergence between \textit{parts of} the NL query and the generated explanation. This misalignment can cause the validator to misinterpret or ``overlook'' the entailment relationship, as illustrated in the following example,}
    \begin{figure}[ht]
    \vspace{-3mm}
      \begin{tikzpicture}
      \node (table) [inner sep=0pt] {
      \begin{adjustbox}{width=\linewidth}
        \begin{tabular}{l} 
         \\ [-2.0ex]
         \rowcolor{gray!10}\makecell[l]{
         \textbf{NL Query:} \textit{Give the names of countries that are in Europe and} \\ \textit{have a population equal to 80000.}\\\\
         \textbf{Predicted SQL Query:} \\
         {\fontfamily{qcr}\fontsize{9}{10}\selectfont\textcolor{pp}{\textbf{SELECT}} name \textcolor{pp}{\textbf{FROM}} country} \\
         {\fontfamily{qcr}\fontsize{9}{10}\selectfont\textcolor{pp}{\textbf{WHERE}} continent = \textcolor{rd}{\textquotesingle Europe\textquotesingle} \textcolor{pp}{\textbf{AND}} population >= 8000}\\
         \textbf{False-Positive Generated Explanation:}\\
         \textit{The query returns a result set with 1 column (name) and 38 rows, filtered by}\\
         \textit{continent equal to Europe and a population \textcolor{rd}{greater than or equal to 80000}.} \\
         \textit{Among them, the result, for example, country, name Estonia, continent is } \\
         \textit{Europe, the population is 1439200 \textcolor{rd}{greater than or equal to 80000}.} \\\\
         \textbf{Ground-truth SQL Query:} \\
         {\fontfamily{qcr}\fontsize{9}{10}\selectfont\textcolor{pp}{\textbf{SELECT}} name \textcolor{pp}{\textbf{FROM}} country} \\
         {\fontfamily{qcr}\fontsize{9}{10}\selectfont\textcolor{pp}{\textbf{WHERE}} continent = \textcolor{rd}{\textquotesingle Europe\textquotesingle} \textcolor{pp}{\textbf{AND}} population = 8000}\\
         \textbf{Generated Explanation for Ground-truth SQL:}\\
         \textit{The query returns a result set with 1 column (name) and 38 rows, filtered by}\\
         \textit{continent equal to Europe and a population greater than or equal to 80000.} \\
         \textit{Among them, the result, for example, country, name Estonia, continent is } \\
         \textit{Europe, the population is 1439200 equal to 80000.}
         }
      \end{tabular}
      \end{adjustbox}
      };
      \draw[line width=0.30mm, rounded corners=.1em](table.north west) rectangle (table.south east);
      \end{tikzpicture}
    \end{figure}
    \vspace{-3mm}
    
    \noindent {Such failures may be mitigated if a more fine-grained supervision signal is provided during validator model training.}

    \item {\textbf{Explanation Quality Problem.} Another source of translation errors stems from the quality of the generated NL explanation, which may convey inaccurate or incorrect semantics. An illustrative example is provided below,}

    \setul{}{1pt}
    \begin{figure}[ht]
      \vspace{-3mm}
      \begin{tikzpicture}
      \node (table) [inner sep=0pt] {
      \begin{adjustbox}{width=\linewidth}
        \begin{tabular}{l} 
         \\ [-2.0ex]
         \rowcolor{gray!10}\makecell[l]{
         \textbf{NL Query:} \textit{Count the number of friends Kyle has.} \\\\
         \textbf{Predicted SQL Query:} \\
         {\fontfamily{qcr}\fontsize{9}{10}\selectfont\textcolor{pp}{\textbf{SELECT}} \textcolor{blue}{count}(*) \textcolor{pp}{\textbf{FROM}} highschooler \textcolor{pp}{\textbf{AS}} T1}\\
         {\fontfamily{qcr}\fontsize{9}{10}\selectfont\textcolor{pp}{\textbf{JOIN}} friend \textcolor{pp}{\textbf{AS}} T2 \textcolor{pp}{\textbf{ON}} \ul{T1.id = T2.friend\_id}}\\
         {\fontfamily{qcr}\fontsize{9}{10}\selectfont\textcolor{pp}{\textbf{WHERE}} T1.name = \textcolor{rd}{\textquotesingle Kyle\textquotesingle}}\\
         \textbf{Ground-truth SQL Query:} \\
         {\fontfamily{qcr}\fontsize{9}{10}\selectfont\textcolor{pp}{\textbf{SELECT}} \textcolor{blue}{count}(*) \textcolor{pp}{\textbf{FROM}} highschooler \textcolor{pp}{\textbf{AS}} T1}\\
         {\fontfamily{qcr}\fontsize{9}{10}\selectfont\textcolor{pp}{\textbf{JOIN}} friend \textcolor{pp}{\textbf{AS}} T2 \textcolor{pp}{\textbf{ON}} \ul{T1.id = T2.student\_id}}\\
         {\fontfamily{qcr}\fontsize{9}{10}\selectfont\textcolor{pp}{\textbf{WHERE}} T1.name = \textcolor{rd}{\textquotesingle Kyle\textquotesingle}}\\\\
         \textbf{Generated Explanations for both SQL Queries:}\\
         \textit{The query returns a result set with one column with aggregation type (count)}\\
         \textit{and 1 row, filtered by name equal to Kyle. The result, name Kyle, there are 2}\\
         \textit{high schoolers with friends, the number of friends is 2.}
         }
      \end{tabular}
      \end{adjustbox}
      };
      \draw[line width=0.30mm, rounded corners=.1em](table.north west) rectangle (table.south east);
      \end{tikzpicture}
    \end{figure}
    \vspace{-3mm}

    \noindent {As shown, \textsc{Cyclesql} fails to interpret the semantics of the join conditions, generating the same explanation for both queries, which leads to incorrect validation of the predicted result. As a result, it is imperative to establish a more reliable explanation generation process for \textsc{Cyclesql}.}




\end{itemize}

\begin{table*}
    \centering
    \captionsetup{font=small}
    \caption{NL queries, SQLs, to-explained results, and corresponding NL explanations produced by \textsc{Cyclesql}.}
    \label{tab: case study}
    \vspace{-2mm}
    \begin{adjustbox}{width=\textwidth}
        \begin{tabular}{|l|l|c|l|} \hline 
         \makecell[c]{\textbf{Natural Language Queries}} &  \makecell[c]{\textbf{Executed SQL Queries}} &  \textbf{To-explained Query Results}& \makecell[c]{\textbf{Natural Language Explanations}}\\ \hline 
        \makecell[l]{\ul{$Q_{1}$}: What is the total number of \\ languages used in Aruba?} & \makecell[l]{{\fontfamily{pcr}\fontsize{9}{10}\selectfont\textcolor{pp}{\textbf{SELECT}} \textcolor{blue}{count}(T2.Language)} \\ {\fontfamily{pcr}\fontsize{9}{10}\selectfont\textcolor{pp}{\textbf{FROM}} Country \textcolor{pp}{\textbf{AS} T1} \textcolor{pp}{\textbf{JOIN}} Countrylanguage \textcolor{pp}{\textbf{AS}} T2} \\ {\fontfamily{pcr}\fontsize{9}{10}\selectfont\textcolor{pp}{\textbf{WHERE}} T1.name = \textcolor{rd}{\textquotesingle Aruba\textquotesingle}}}  & \begin{tabular}{|c|} \hline
              \rowcolor{pale} \textbf{count(T2.language)} \\ \hline
              \rowcolor{white} 4 \\ \hline
         \end{tabular} & \makecell[l]{\textit{The query output is a result set with one column and one } \\ \textit{row, filtered by country name Aruba. In this specific result, } \\ \textit{country Aruba, whose country code is ABW, has four }\\
         \textit{spoken languages. So the count of languages is 4.}} \\ \hline 
         
        \makecell[l]{\ul{$Q_{2}$}: What is the continent name \\ that Anguilla belongs to?} & \makecell[l]{{\fontfamily{pcr}\fontsize{9}{10}\selectfont\textcolor{pp}{\textbf{SELECT}} continent \textcolor{pp}{\textbf{FROM}} Country} \textcolor{pp}{\textbf{WHERE}} name = \textcolor{rd}{\textquotesingle Anguilla\textquotesingle}} & \begin{tabular}{|c|} \hline
              \rowcolor{pale} \textbf{continent} \\ \hline
              \rowcolor{white} North America \\ \hline
         \end{tabular} & \makecell[l]{\textit{The query returns a result set with one column and one row,} \\ \textit{filtered by country name Anguilla. Here, country Anguilla, }\\ \textit{belongs to the continent North America.}} \\ \hline

        \makecell[l]{\ul{$Q_{3}$}: What are the names of nations \\ speak both English and French?} & \makecell[l]{{\fontfamily{pcr}\fontsize{9}{10}\selectfont\textcolor{pp}{\textbf{SELECT}} T1.name \textcolor{pp}{\textbf{FROM}} Country \textcolor{pp}{\textbf{AS}} T1} \\ {\fontfamily{pcr}\fontsize{9}{10}\selectfont\textcolor{pp}{\textbf{JOIN}} Countrylanguage \textcolor{pp}{\textbf{AS}} T2 \textcolor{pp}{\textbf{ON}} T1.code = T2.countrycode} \\ {\fontfamily{pcr}\fontsize{9}{10}\selectfont\textcolor{pp}{\textbf{WHERE}} T2.language = \textcolor{rd}{\textquotesingle English\textquotesingle} \textcolor{pp}{\textbf{INTERSECT}}} \\ {\fontfamily{pcr}\fontsize{9}{10}\selectfont\textcolor{pp}{\textbf{SELECT}} T1.name \textcolor{pp}{\textbf{FROM}} Country \textcolor{pp}{\textbf{AS}} T1} \\ {\fontfamily{pcr}\fontsize{9}{10}\selectfont\textcolor{pp}{\textbf{JOIN}} Countrylanguage \textcolor{pp}{\textbf{AS}} T2 \textcolor{pp}{\textbf{ON}} T1.code = T2.countrycode} \\ {\fontfamily{pcr}\fontsize{9}{10}\selectfont\textcolor{pp}{\textbf{WHERE}} T2.language = \textcolor{rd}{\textquotesingle French\textquotesingle}}} & \begin{tabular}{|c|} \hline
              \rowcolor{pale} \textbf{T1.name} \\ \hline
              \rowcolor{white} Seychelles \\ \hline
         \end{tabular} & \makecell[l]{\textit{The query output is a result set with one column and 6 rows,} \\ \textit{filtered by language English or French. Among them, for} \\ \textit{example, country Seychelles, where its country languages} \\ \textit{include English and French.}} \\ \hline 
         
         \makecell[l]{\ul{$Q_{4}$}: Which cities are in European \\ countries where English is not \\ the official language?} & \makecell[l]{{\fontfamily{pcr}\fontsize{9}{10}\selectfont\textcolor{pp}{\textbf{SELECT}} \textcolor{pp}{\textbf{DISTINCT}} T2.name \textcolor{pp}{\textbf{FROM}} Country \textcolor{pp}{\textbf{AS}} T1} \\ {\fontfamily{pcr}\fontsize{9}{10}\selectfont\textcolor{pp}{\textbf{JOIN}} City \textcolor{pp}{\textbf{AS}} T2 \textcolor{pp}{\textbf{ON}} T1.code = T2.countrycode }
     \\ {\fontfamily{pcr}\fontsize{9}{10}\selectfont\textcolor{pp}{\textbf{WHERE}} T1.Continent  =  \textcolor{rd}{\textquotesingle Europe\textquotesingle} \textcolor{pp}{\textbf{AND}} T1.Name \textcolor{pp}{\textbf{NOT IN}} (} \\ {\hspace{0.2cm} \fontfamily{pcr}\fontsize{9}{10}\selectfont\textcolor{pp}{\textbf{SELECT}} T3.name \textcolor{pp}{\textbf{FROM}} Country \textcolor{pp}{\textbf{AS}} T3} \\
     {\hspace{0.2cm} \fontfamily{pcr}\fontsize{9}{10}\selectfont\textcolor{pp}{\textbf{JOIN}} Countrylanguage \textcolor{pp}{\textbf{AS}} T4 \textcolor{pp}{\textbf{ON}} T3.code = T4.countrycode} \\ {\hspace{0.2cm} \fontfamily{pcr}\fontsize{9}{10}\selectfont\textcolor{pp}{\textbf{WHERE}} T4.isofficial = \textcolor{rd}{\textquotesingle T\textquotesingle} \textcolor{pp}{\textbf{AND}} T4.language = \textcolor{rd}{\textquotesingle English\textquotesingle})}}  & \begin{tabular}{|c|} \hline
              \rowcolor{pale} \textbf{T2.name} \\ \hline
              \rowcolor{white} Nabereznyje Tšelny \\ \hline
         \end{tabular} & \makecell[l]{\textit{The query output is a result set with one column comprising} \\ \textit{753 rows, filtered by continent Europe and excludes entries} \\ \textit{that official is not T and language is English. In this specific} \\ \textit{result, for example, the city Nabereznyje Tšelny in the} \\ \textit{Russian Federation, situated in Europe and not meeting the} \\ \textit{specified language and official status criteria.}}\\ \hline 
         
        \makecell[l]{\ul{$Q_{5}$}: Return the country name and \\ the numbers of languages spoken \\ for each country that speaks at \\ least 3 languages.} & \makecell[l]{{\fontfamily{pcr}\fontsize{9}{10}\selectfont\textcolor{pp}{\textbf{SELECT}} \textcolor{blue}{count}(T2.language), T1.name \textcolor{pp}{\textbf{FROM}} Country \textcolor{pp}{\textbf{AS}} T1} \\ {\fontfamily{pcr}\fontsize{9}{10}\selectfont\textcolor{pp}{\textbf{JOIN}} Countrylanguage \textcolor{pp}{\textbf{AS}} T2 \textcolor{pp}{\textbf{ON}} T1.code = T2.countrycode}
     \\ {\fontfamily{pcr}\fontsize{9}{10}\selectfont\textcolor{pp}{\textbf{GROUP BY}} T1.name \textcolor{pp}{\textbf{HAVING}} \textcolor{blue}{count}(*) > 2}} & \begin{tabular}{|c|c|} \hline
              \rowcolor{pale} \textbf{count(T2.language)} & \textbf{T1.name} \\ \hline
              \rowcolor{white} 5 & Iraq \\ \hline
         \end{tabular} & \makecell[l]{\textit{The query result has two columns with 149 rows, filtered by} \\ \textit{country language greater than 2. Among these results, 149} \\ \textit{entries have 2 more languages, and for the specific example} \\ \textit{provided, the country Iraq has 5 spoken languages.}}\\ \hline 
        \end{tabular}
    \end{adjustbox}
\end{table*}


\subsection{Qualitative Evaluation}\label{5.2}
The NL explanation introduced in \textsc{Cyclesql} is designed to rationalize the query result and hence provides meaningful information to understand the overall NL2SQL translation. An inherent query may be raised: \textit{Can the semantics expressed in the NL explanation help users understand existing black-box NL2SQL translation?} Therefore, we further evaluate the quality of explanations produced by \textsc{Cyclesql} using a case study and a user study on the database (named \textit{world\_1}) in \textsc{Spider} benchmark to answer the question.

Note that in order to enhance the fluency of NL explanations for users, we choose to use a ``polishing model'' to help refine the explanations further. Despite many studies on text deliberation \cite{deliberation16, deliberation17}, it is noteworthy that recent emergent LLMs have showcased superior text deliberation capabilities \cite{GPT3-20, instruction22}. Hence, we simply use the $5$-shot prompted \textsc{Gpt-3.5-turbo} LLM (with simple human-written instructions) as the ``polishing model'' to enhance the NL explanations and use the refined explanations for the subsequent evaluations.



\subsubsection{Case Study.} The \textit{world\_1} dataset is an SQLite database provided by \textsc{Spider}. It provides an overview of the global city and country data, facilitating queries on various aspects of worldwide demographics and linguistic diversity. The database contains $4$ relational tables and about $5000$ data rows.

\noindent\textbf{Setup.} Since the \textsc{Spider} benchmark includes 120 NL-SQL pairs for the \textit{world\_1} database, we selected five queries from the benchmark for this evaluation. To show NL explanations generated by \textsc{Cyclesql} for different types of queries, we simply cluster the pairs based on the structures of SQL queries and randomly choose five query pairs from each cluster. Table~\ref{tab: case study} shows the \textit{NL queries}, \textit{executed SQL queries}, \textit{to-explained query results}, and the \textit{NL explanations} returned by \textsc{Cyclesql} for these selected queries.

\noindent\textbf{Analysis.} We analyze each example as follows:
\begin{itemize}[leftmargin=*]
\item $Q_{1}$: This question inquires about the total number of spoken languages in the country Aruba. The explanation first describes the result set and then explains that for the country, Aruba, the total number of its spoken languages is $4$. 
\item $Q_{2}$: Retrieve the continent name with the country Anguilla. The NL explanation is simple and concise, which tells the country Anguilla should belong to North America. Note that for those simple queries, the NL explanation generated by \textsc{Cyclesql} may be similar to other explanation methods, such as the SQL2NL method we introduced before.
\item $Q_{3}$: This NL query asks to find the nation name that satisfies its condition of ``\textit{speak both English and French}''. While the underlying executed SQL query is complex, involving an {\fontfamily{pcr}\fontsize{9}{10}\selectfont\textcolor{pp}{\textbf{INTERSECT}}} set operation that combines two queries, the NL explanation is straightforward and precisely elucidates why Seychelles is one of the query results. 
\item $Q_{4}$: The negation question asks for European cities that do not use English as their official language. Although the NL explanation initially presents the filtering conditions somewhat incoherently, it summarizes that 753 countries meet these criteria. Following this, the explanation specifies that the city Nabereznyje Tšelny is located in Europe and fails to meet the language and official criteria. Despite the obscurity surrounding the ``\textit{English}'' filtering condition in the NL explanation, it still offers valuable insights to help users comprehend the query result.
\item $Q_{5}$: The question aims to identify countries with at least 3 spoken languages. The explanation centers on the result of the country Iraq. It first provides an overall total number of the countries that speak two or more languages (i.e., 149 countries). Then, the explanation shifts its focus to the country Iraq, stating that Iraq speaks 5 languages in total.

\end{itemize}

\subsubsection{User Study.} Next, we report a user study to assess the quality of the explanations generated by \textsc{Cyclesql} for the five queries selected from the \textit{world\_1} database. We focus on addressing the main inquiry: \textit{whether \textsc{Cyclesql} provides meaningful explanations for users to understand query results, hence the underlying translations.}

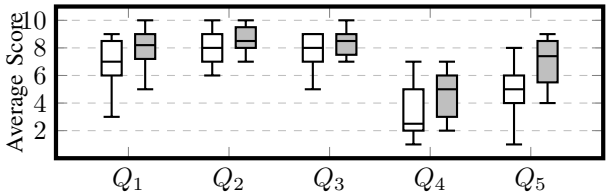
\begin{figure}[ht]
\vspace{-3mm}
\begin{subfigure}[b]{1.0\linewidth}
  \centering
  \caption{Summarized results on NL explanations by users.}
  \begin{adjustbox}{width=\linewidth}
  \begin{tabular}{l c c c c c c c c c c} \hline
    \multirow{2}{*}{\textbf{Dimension}} & \multicolumn{5}{|c}{\textbf{\textsc{Gpt-3.5-turbo}}} & \multicolumn{5}{|c}{\textbf{\textsc{Cyclesql}}} \\ 
    & \multicolumn{1}{|c}{$Q_{1}$} & $Q_{2}$ & $Q_{3}$ & $Q_{4}$ & $Q_{5}$ & \multicolumn{1}{|c}{$Q_{1}$} & $Q_{2}$ & $Q_{3}$ & $Q_{4}$ & $Q_{5}$ \\\Xhline{3\arrayrulewidth}
    \textit{Query result interpretability}  & \multicolumn{1}{|c}{\twemoji[height=1.0em]{face with rolling eyes}} & \twemoji[height=1.0em]{party popper} & \twemoji[height=1.0em]{party popper} & \twemoji[height=1.0em]{nauseated face} & \twemoji[height=1.0em]{nauseated face} & \multicolumn{1}{|c}{\twemoji[height=1.0em]{party popper}} & \twemoji[height=1.0em]{party popper} & \twemoji[height=1.0em]{party popper} & \twemoji[height=1.0em]{face with rolling eyes} & \twemoji[height=1.0em]{face with rolling eyes} \\ \hline
    \textit{Textual entailment with NL}  & \multicolumn{1}{|c}{\twemoji[height=1.0em]{party popper}} & \twemoji[height=1.0em]{party popper} & \twemoji[height=1.0em]{party popper} & \twemoji[height=1.0em]{face with rolling eyes} & \twemoji[height=1.0em]{face with rolling eyes} & \multicolumn{1}{|c}{\twemoji[height=1.0em]{party popper}} & \twemoji[height=1.0em]{party popper} & \twemoji[height=1.0em]{party popper} & \twemoji[height=1.0em]{face with rolling eyes} & \twemoji[height=1.0em]{party popper} \\ \Xhline{3\arrayrulewidth}
    Overall Ratings  & \multicolumn{1}{|c}{\twemoji[height=1.0em]{face with rolling eyes}} & \twemoji[height=1.0em]{party popper} & \twemoji[height=1.0em]{party popper} & \twemoji[height=1.0em]{nauseated face} & \twemoji[height=1.0em]{face with rolling eyes} & \multicolumn{1}{|c}{\twemoji[height=1.0em]{party popper}} & \twemoji[height=1.0em]{party popper} & \twemoji[height=1.0em]{party popper} & \twemoji[height=1.0em]{face with rolling eyes} & \twemoji[height=1.0em]{party popper}  \\\hline
  \end{tabular}
  \end{adjustbox}
  \label{tab: user study emoji results}
\end{subfigure}
\par\bigskip
\begin{subfigure}[b]{1.0\linewidth}
    \centering
    \definecolor{airforceblue}{rgb}{0.36, 0.54, 0.66}
    \definecolor{antiquebrass}{rgb}{0.8, 0.58, 0.46}
    \definecolor{darkbrown}{rgb}{0.4, 0.26, 0.13}
    \definecolor{wildblueyonder}{rgb}{0.64, 0.68, 0.82}
    \begin{tikzpicture}
      \begin{axis} [
        boxplot/draw direction = y,
        height=3.6cm, 
        width=\linewidth,
        ymin = 0,
        ymax = 11,
        ytick={2,4,...,10},
        axis line style=ultra thick,
        ylabel={Average Score},
        ylabel style={font=\small,yshift=-4.5ex},
        boxplot={
            %
            draw position={1/3 + floor(\plotnumofactualtype/2) + 1/3*mod(\plotnumofactualtype,2)},
            %
            box extend=0.20,
        },
        xticklabel={font=\small},
        xtick={0.5,1.5,2.5,3.5,4.5},
        xticklabels={$Q_{1}$, $Q_{2}$, $Q_{3}$, $Q_{4}$, $Q_{5}$}, 
        ymajorgrids=true,
        grid style=dashed,
        boxplot/every box/.style={draw=black,solid,thick,fill=gray!50},
        boxplot/every whisker/.style={color=black,solid,thick},
        boxplot/every median/.style={color=black,solid,thick},
        ]
     
        \addplot+ [
        boxplot prepared={
          upper quartile=8.5,
          lower quartile=6,
          upper whisker=9,
          lower whisker=3,
          median=7.0,
          every box/.style={draw=black,thick,fill=white}
        }
        ] coordinates {};
        \addplot+ [
        boxplot prepared={
          upper quartile=9,
          lower quartile=7.2,
          upper whisker=10,
          lower whisker=5,
          median=8.2,
          every box/.style={draw=black,thick,fill=gray!50}
        }
        ] coordinates {};

        \addplot+ [
        boxplot prepared={
          upper quartile=9,
          lower quartile=7,
          upper whisker=10,
          lower whisker=6,
          median=8,
          every box/.style={draw=black,thick,fill=white}
        },
        ] coordinates {};
        \addplot+ [
        boxplot prepared={
          upper quartile=9.5,
          lower quartile=8,
          upper whisker=10,
          lower whisker=7,
          median=8.5,
          every box/.style={draw=black,thick,fill=gray!50}
        }
        ] coordinates {};

        \addplot+ [
        boxplot prepared={
          upper quartile=9,
          lower quartile=7,
          upper whisker=9,
          lower whisker=5,
          median=8,
          every box/.style={draw=black,thick,fill=white}
        },
        ] coordinates {};
        \addplot+ [
        boxplot prepared={
          upper quartile=9,
          lower quartile=7.5,
          upper whisker=10,
          lower whisker=7,
          median=8.5,
          every box/.style={draw=black,solid,thick,fill=gray!50}
        }
        ] coordinates {};

        \addplot+ [
        boxplot prepared={
          upper quartile=5,
          lower quartile=2,
          upper whisker=7,
          lower whisker=1,
          median=2.5,
          every box/.style={draw=black,solid,thick,fill=white}
        }
        ] coordinates {};
        \addplot+ [
        boxplot prepared={
          upper quartile=6,
          lower quartile=3,
          upper whisker=7,
          lower whisker=2,
          median=5,
          every box/.style={draw=black,solid,thick,fill=gray!50}
        }
        ] coordinates {};

        \addplot+ [
        boxplot prepared={
          upper quartile=6,
          lower quartile=4,
          upper whisker=8,
          lower whisker=1,
          median=5,
          every box/.style={draw=black,solid,thick,fill=white}
        }
        ] coordinates {};
        \addplot+ [
        boxplot prepared={
          upper quartile=8.5,
          lower quartile=5.5,
          upper whisker=9,
          lower whisker=4,
          median=7.4,
          every box/.style={draw=black,solid,thick,fill=gray!50}
        }
        ] coordinates {};
      \end{axis}
    \end{tikzpicture}
    \vspace{-2mm}
    \caption{Average ratings with white boxes indicating ratings for explanations by \textsc{Gpt-3.5-turbo} and gray boxes for \textsc{Cyclesql}.}
    \label{tab: averaging rating results}
\end{subfigure}
\vspace{-5mm}
\captionsetup{font=small}
\caption{User study results}
\label{fig: user study results}
\end{figure}

\noindent\textbf{Participants and Tasks.} We enlisted the participation of 20 computer science students who have prior knowledge of SQL or have taken graduate-level database courses to join the assessment procedure. Initially, participants were briefed about the background of the study and showed the schema of the database. We then provided an exemplary query with a ``perfect'' explanation (human-written) to illustrate the evaluation process. Participants were then tasked to assign scores, ranging from $1$ to $10$, to each explanation for every query. The explanations were provided either by \textsc{Cyclesql} or the simpler method with \textsc{Gpt-3.5-turbo} as we described in Section~\ref{5.1.4}. Participants are mainly considered in two aspects (refer to Table~\ref{tab: user study emoji results}) and provided an overall rating. To simplify the analysis, we computed average scores for each query and also summarized the results (in the form of \twemoji[height=1.0em]{party popper}, \twemoji[height=1.0em]{face with rolling eyes}, and \twemoji[height=1.0em]{nauseated face}). Here, \twemoji[height=1.0em]{party popper} denotes ``great'' for scores ranging from $7$ to $10$, \twemoji[height=1.0em]{face with rolling eyes} denotes ``neutral'' representing scores ranging from $3$ to $7$, and \twemoji[height=1.0em]{nauseated face} denotes ``bad'' for scores between $0$ and $3$. Fig.~\ref{fig: user study results} shows the results for this user study.

\noindent\textbf{Results and Analysis.} The feedback is positive for \textsc{Cyclesql}: $14$ out of $20$ participants agreed that the explanations generated by \textsc{Cyclesql} are more comprehensible to them compared with the ones from \textsc{Gpt-3.5-turbo}. Moreover, the average rating scores consistently surpass those of \textsc{Gpt-3.5-turbo}, as illustrated in Fig.~\ref{tab: averaging rating results}. More importantly, participants confirmed that examining these explanations can greatly assist them in understanding the reasons behind the query results and may enable them to detect potential translation errors that happened with the underlying NL2SQL process.

\noindent\textbf{Discussion.} While \textsc{Cyclesql} provides an autonomous feedback loop in its current form, the findings suggest that the NL explanation can also serve as feedback for humans when interacting with NLIDBs, facilitating real-time adaptions based on human inputs. In future work, we intend to further explore the human feedback loop in the context of NL2SQL.

\section{Related Work}\label{section 6}
\noindent\textbf{Natural Language Interface for Database.} NLIDBs have been studied both in database management and NLP communities. Many machine learning-based approaches treat the NL2SQL problem as a Seq2seq translation task and leverage the encoder-decoder architecture to do the translation \cite{Zhong17, Yu18, Bogin19, IRNet19, RATSQL20, GraPPa20, BRIDGE20, smbop21, LGESQL20, resdsql23, Graphix23, CatSQL23, MIGA22, gary24}. With the excellent performance of LLMs in various NLP tasks, recent works have explored applying LLMs to the NL2SQL task \cite{nitarshan22, aiwei23, DINSQL23, hybrid23, SQL-PaLM23, purple24}, mostly in an end-to-end manner.



Instead of generating the target queries end-to-end, \textsc{Cyclesql} aims to create a feedback loop within the translation process, enabling the potential to find a better SQL outcome.

\noindent\textbf{Explanations for Query Answers.} This line of research aims to explain query outcomes, including missing tuples, outlier values, or returned tuples. Provenance provides a direct method of explanation by characterizing a set of tuples contributing to the target query answer \cite{whynot09, BidoitHT14, Deutch17, Deutch20, missing20, MiaoZGR19, CaJaDE21}. Prior works \cite{whynot09, BidoitHT14} have explored both ``why'' and ``why-not'' questions through query-based explanation methods that make changes to queries. The frameworks proposed by \cite{Scorpion13, RoyS14, RoyOS15} address explanations for outlier values, particularly in aggregate query results, which present explanations as compact summaries of relevant tuples. Conversely, \cite{MiaoZGR19, CaJaDE21} delve into explanations for outliers beyond provenance, exploring learning patterns or augmenting provenance. 

Explanations for returned tuples have been studied in \cite{Deutch17, Deutch20}, with an emphasis on formulating the provenance of output tuples in NL, either in factorized or summarized form. We align with their motivations in considering explanations in NL, but step further to utilize these explanations back to the NL2SQL translation process, thereby providing self-generated feedback to enhance the underlying translation accuracy.

\section{Conclusion \& Future Work}\label{section 7}
 In this paper, we proposed an iterative framework named \textsc{Cyclesql} for end-to-end models to autonomously generate the best output using NL explanations. \textsc{Cyclesql} initials by tracking data provenance and query semantics for a given query result to generate an NL explanation. The explanation is then used to validate the correctness of the underlying translation, ultimately resulting in a better translation outcome. We conducted extensive experiments on five benchmarks, showing that \textsc{Cyclesql} consistently improves translation performance across seven models. Additional case and user studies provided further insights into the quality of the generated explanations.

 While \textsc{Cyclesql} has proven effective, further efforts are needed to enhance its capabilities. One potential area of investigation is how to close the feedback loop with human involvement, enabling \textsc{Cyclesql} to better align with humans and adapt its translation over time. In addition, developing a more reliable way to assist data tracking in \textsc{Cyclesql} is critical for \textsc{Cyclesql}, especially in better supporting empty-result queries. Finally, an intended future research direction involves exploring an effective way to incorporate fine-grained semantics both in NL queries and explanations during the training of the NLI model, thereby enhancing its resilience in managing comparable translations.

\section{Acknowledgments}
The authors would like to thank all the anonymous reviewers for their insightful comments and suggestions. This work was supported by NSFC (Grant No. 62272106).

\balance
\bibliographystyle{IEEEtran}
\bibliography{conference_101719}

\end{document}